\documentclass[10pt,journal,compsoc]{IEEEtran}
\ifCLASSOPTIONcompsoc
  \usepackage[nocompress]{cite}
  \usepackage{pdfpages}
  \usepackage{placeins}
  \usepackage{pifont}
  \usepackage{utfsym}
  \usepackage{multirow}
  \usepackage{threeparttable}
  \usepackage{makecell}
  \usepackage{graphicx}
  \usepackage{array}
  \usepackage{amssymb}
  \usepackage{amsmath}
\else
  \usepackage{cite}
\fi
\ifCLASSINFOpdf
\else
\fi
\begin{document}
\title{Temporal Aware Mixed Attention-based Convolution and Transformer Network (MACTN) for EEG Emotion Recognition}

\author{Xiaopeng~Si,~\IEEEmembership{Member, ~IEEE,}
        Dong~Huang,
        Yulin~Sun,
        and~Dong~Ming,~\IEEEmembership{Senior~Member,~IEEE}% <-this % stops a space
\IEEEcompsocitemizethanks{
% \IEEEcompsocthanksitem M. Shell was with the Department
% of Electrical and Computer Engineering, Georgia Institute of Technology, Atlanta,
% GA, 30332.\protect\\
% % note need leading \protect in front of \\ to get a newline within \thanks as
% % \\ is fragile and will error, could use \hfil\break instead.
% E-mail: see http://www.michaelshell.org/contact.html
% \IEEEcompsocthanksitem J. Doe and J. Doe are with Anonymous University.
% \IEEEcompsocthanksitem $^1$Academy of Medical Engineering and Translational Medicine, Tianjin University, Tianjin 300072, China;
% \IEEEcompsocthanksitem $^2$Tianjin Key Laboratory of Brain Science and Neural Engineering, Tianjin University, Tianjin 300072, China;
\IEEEcompsocthanksitem X. Si and D. Huang contributed equally to this work.
\IEEEcompsocthanksitem X. Si, D. Huang, Y. Sun, and D. Ming are with the Academy of Medical Engineering and Translational Medicine, Tianjin University, Tianjin
300072, China, and also with Tianjin Key Laboratory of Brain Science
and Neural Engineering, Tianjin University, Tianjin 300072, China (E-mail:\{xiaopengsi, huang\_dong, syuri, richardming\}@tju.edu.cn).
\IEEEcompsocthanksitem X. Si and D. Ming are the corresponding authors.

% \IEEEcompsocthanksitem \IEEEauthorrefmark{1}To whom correspondence should be addressed.
% \IEEEcompsocthanksitem \IEEEauthorrefmark{1}Corresponding author: Dong Ming (richardming@tju.edu.cn)
}% <-this % stops an unwanted space
% \thanks{Manuscript received November 30, 2022; revised August 26, 2015.}
}


\markboth{Journal of \LaTeX\ Class Files,~Vol.~14, No.~8, November~2022}%
{Shell \MakeLowercase{\textit{et al.}}: Bare Demo of IEEEtran.cls for Computer Society Journals}



\IEEEtitleabstractindextext{%
\iffalse
% 【对应引言】情绪识别在人机交互中起着重要的作用，脑电图有利于反映人类的情绪状态。然而，人类真实生活场景中较多的基本情绪以及不同人的情绪体验差异比较大，对基于EEG的情绪识别的实际应用提出了巨大挑战。
% 【对应方法】受到神经科学中有关情绪的时间动态性研究的启发，情绪的时间动态性指，情绪在持续时间上高度可变的，由此我们提出一个充分考虑情绪时间动态性的、端到端、不受被试个体差异影响的方法，该方法分两步提取时间上的情绪特征，第一步通过卷积神经网络（CNN）提取局部情绪特征，第二步通过transformer整合稀疏的全局情感特征，其中穿插有通道注意力机制提取通道维度上特征。
% 【对应结果】该方法在情绪档案（THU-EP）数据集上获得了58.6±5.7%和58.7±5.9%的9分类准确率和F1 score，准确率超过现有最先进的方法11.6%，并且该方法的相关变体获得了2022年世界机器人大赛——脑控机器人大赛情绪赛题的冠军。一般的，端到端的方法具有较低的可解释性，我们通过4种方法对我们的模型做出了解释，这些方法证明了提出的模型的有效性，刺激材料情绪强度较高与模型中自注意力权重较高的时间段的对应，显示了提出的模型在时间维度上进行情绪定位方面的潜力。
\fi
\begin{abstract}
Emotion recognition plays a crucial role in human-computer interaction, and electroencephalography (EEG) is advantageous for reflecting human emotional states. In this study, we propose MACTN, a hierarchical hybrid model for jointly modeling local and global temporal information. The model is inspired by neuroscience research on the temporal dynamics of emotions. MACTN extracts local emotional features through a convolutional neural network (CNN) and integrates sparse global emotional features through a transformer. Moreover, we employ channel attention mechanisms to identify the most task-relevant channels. Through extensive experimentation on two publicly available datasets, namely THU-EP and DEAP, our proposed method, MACTN, consistently achieves superior classification accuracy and F1 scores compared to other existing methods in most experimental settings. Furthermore, ablation studies have shown that the integration of both self-attention mechanisms and channel attention mechanisms leads to improved classification performance. Finally, an earlier version of this method, which shares the same ideas, won the Emotional BCI Competition's final championship in the 2022 World Robot Contest.

\end{abstract}

}

\maketitle

\IEEEdisplaynontitleabstractindextext

% \fi
\IEEEpeerreviewmaketitle

\IEEEraisesectionheading{\section{Introduction}\label{sec:introduction}}

% for IEEE Computer Society journal papers produced under \LaTeX\ using
% IEEEtran.cls version 1.8b and later.
% % You must have at least 2 lines in the paragraph with the drop letter
% % (should never be an issue)
% I wish you the best of success.
\iffalse
para 1-2 intro
\fi
\IEEEPARstart{E}{motions} are crucial for human daily life, as they directly influence people’s judgment, memory, behavior and social interaction \cite{mumenthaler2018emotion}. Emotion recognition is essential for human-computer interaction, as it enables machines to perceive human affective states and make them more “empathetic” in their interactions with humans. Negative emotions may lead to serious brain disorders such as mental illness \cite{vos2017global}, and emotion recognition can also facilitate doctors’ assessment of psychological health and disorders such as autism \cite{kuusikko2009emotion}. Furthermore, emotion recognition plays a key role in cognitive behavioral therapy \cite{carpenter2018cognitive}, emotion regulation therapy/emotion-focused therapy \cite{lane2015memory, huang2021neurofeedback}.
Facial expressions, language, gestures, physiological signals and other modalities are often used for emotion recognition. Based on the type of data, emotion recognition methods can be divided into two categories: one is based on non-physiological signals, such as facial expression images, body posture and voice signals; the other is based on physiological signals, such as electroencephalogram (EEG), electromyogram (EMG), electrocardiogram (ECG). Among various types of physiological signals, EEG signal is one of the most commonly used ones. It captures directly from the cerebral cortex, thus it is advantageous for reflecting human mental states. EEG also has high temporal resolution, non-invasiveness and high sensitivity to external stimuli \cite{van2009brain}.

\iffalse
para 3 intro
\fi
In recent years, EEG-based emotion recognition has attracted wide attention from researchers. Duan et al. \cite{duan2013differential} used differential entropy (DE) features of EEG and support vector machine (SVM) to recognize emotions. Chen et al. \cite{chen2021personal} proposed a Personal-Zscore feature processing method to process the extracted statistical features, time-frequency features and so on, which improved the accuracy of SVM for emotion recognition. Recent studies have shown that deep learning has achieved good results in EEG classification tasks, such as motor imagery \cite{kwon2019subject, tabar2016novel}, emotion recognition \cite{yang2017eeg, li2018hierarchical}, epilepsy detection \cite{sun2022continuous, li2022seizure}, sleep staging classification \cite{khalili2021automatic, sohaib2022automated} and so on. Yang et al. \cite{yang2017eeg} designed a hierarchical network structure and used key subnetwork nodes to improve the performance of emotion recognition. Li et al. \cite{li2018hierarchical} mapped the DE features extracted from EEG to a two-dimensional brain space and used a hierarchical convolutional neural network (HCNN) to extract the emotional representation of EEG in the two-dimensional space. Although many machine learning methods have been proposed for emotion recognition, most of them rely heavily on handcrafted features.

\iffalse
para 4 intro
\fi
Convolutional neural networks (CNNs) and attention-based Transformers have the ability to learn features directly from samples. CNNs have strong local feature extraction capabilities and achieve good results in EEG classification tasks. Schirrmeister et al. \cite{schirrmeister2017deep} proposed deep convolutional neural networks and shallow convolutional neural networks (DeepConvNet and shallow convnet) to process EEG data and obtained better classification performance. Lawhyn et al. \cite{lawhern2018eegnet} proposed EEGNet, which extracts spatial and temporal features by using one-dimensional convolutional kernels on different scales. Transformers can focus on sparse global features and process sequential data such as EEG in parallel, thus attracting the attention of BCI researchers. Moreover, studies have shown that emotions are dynamic, changing in intensity, duration and other attributes \cite{hakim2013computational}, and their duration is highly variable, ranging from a few seconds to several hours or even longer \cite{verduyn2015determinants}. The Transformer model based on self-attention mechanism can fully consider the temporal dynamics of emotions by assigning different weights to sampling points/segments on the time dimension. Wei et al. \cite{wei2023tc} transformed EEG into time-frequency domain by wavelet transform and recognized it by capsule (or windowed) Transformer. Peng et al. \cite{peng2023temporal} mapped EEG to two-dimensional space and used Transformer for classification after word encoding of two-dimensional space EEG. Considering that CNNs and Transformers have the ability to extract features at different scales, Sun et al. \cite{sun2022multi} mapped EEG to two-dimensional space and combined 3D CNN with Transformer for emotion recognition. And EEG contains rich spatial information (EEG channels correspond to different brain regions), channel attention as a part of attention mechanism has received researchers’ attention. Zhang et al.\cite{zhang2022eeg} used cascaded self-attention to extract emotional information on the time dimension hierarchically, using Squeeze-and-Excitation(SE) channel attention mechanism to select task-related channels. Tao et al.\cite{tao2020eeg} proposed an attention-based convolutional recurrent neural network that combines SE channel attention with self-attention to extract more features to improve emotion recognition performance.  Although many methods for emotion recognition combining CNNs with Transformers have been proposed, most of them are focused on subject-dependent random-shuffle emotion recognition rather than the more challenging cross-subject and cross-session emotion recognition scenarios \cite{chen2021ms}.

\iffalse
para 5-6 intro, remove para 6
\fi
Generally, EEG-based emotion recognition can be divided into subject-dependent emotion recognition and cross-subject emotion recognition. The former relies on building individual classification models for each subject, while the latter relies on the generalization of the model, eliminating the need to build independent classification models for each subject. Hu et al. \cite{hu2022similar} suggested that there are individual differences in emotions. Although cross-subject emotion recognition has lower accuracy than subject-dependent emotion recognition, it has gained widespread attention due to its higher practicality. Song et al. \cite{song2018eeg} proposed a new dynamic graph convolutional neural network (DGCNN) model for emotion recognition, which models multi-channel EEG features using a graph. Shen et al. \cite{shen2022contrastive} proposed a model called CLISA, which constructs a loss function based on contrastive learning and trains the model using a combination of time and spatial convolutions. The model achieved an accuracy of 47\% on the THU-EP dataset. Since EEG signals from the same subject have non-stationarity across different time periods \cite{sanei2013eeg}, and current emotion induction experiments mainly use video induction, subject-dependent emotion recognition can be divided into cross-trial emotion recognition and random shuffling emotion recognition (random shuffling can also be used on the entire dataset). Ding et al. \cite{ding2022tsception} proposed a multi-scale temporal and spatial convolutional neural network called TSception to capture the temporal dynamics and spatial asymmetry of EEG for cross-trial emotion recognition. Furthermore, researchers are actively exploring the interpretability of deep learning methods. Lawhern et al. \cite{lawhern2018eegnet} have granted the model a certain level of interpretability by visualizing temporal and spatial convolutions. In the aforementioned work, CLISA and TSception also endeavored to offer explanations for the proposed models. However, there are few studies that provide interpretability for methods based on attention mechanisms.

\iffalse
para 7-9 intro
\fi
To address the above issues, this paper proposes a model called Mixed Attention based Convolution and Transformer Network (MACTN) that combines CNN and Transformer for cross-trial and cross-subject emotion recognition. EEG signals are directly fed into MACTN, which is an end-to-end deep learning method that requires minimal domain knowledge for feature engineering. Inspired by the Vision Transformer \cite{dosovitskiy2020vit}, MACTN extracts local temporal features through 1D convolutions applied to the time dimension. Grouped convolutions map signals from different EEG channels to more feature channels for feature encoding. The Selective Kernel (SK) channel attention module selects channels that are most relevant to the task with different temporal convolution kernels, and the self-attention mechanism is used to extract global sparse emotion features.

Experiments on two publicly available benchmark datasets, Emotion Profile (THU-EP) \cite{hu2022similar} and Database for Emotion Analysis using Physiological signals (DEAP) \cite{koelstra2011deap}, were conducted to evaluate the performance of MACTN. MACTN was compared with several advanced deep and non-deep methods in the BCI field, and in most experiments, MACTN achieved higher accuracy and F1 scores than other methods. Specifically, an earlier version of MACTN with the same concept won the championship in the Brain-Controlled Robot Contest at the 2022 World Robot Contest. A ablation study was conducted to analyze the contribution of each module in MACTN. In addition, we provide some visualization methods to explain the model, such as feature map visualization, convolution kernel visualization, channel attention visualization, and self-attention weight visualization combined with stimuli materials. These methods intuitively illustrate the working principle of the proposed model.

The major contribution of this work can be summarised as:
\begin{itemize}

\item We propose MACTN, a hybrid model based on CNN and Transformer. MACTN extracts local temporal features through convolution and integrates them using channel attention with different kernel sizes. Finally, MACTN uses self-attention mechanism to extract global, sparse emotional features.
\item We conducted extensive experiments on the THU-EP and DEAP datasets. In most of the experiments, MACTN achieved higher accuracy and F1 scores than other methods.
\item Due to the large amount of data in the THU-EP dataset, which includes data from 80 participants, we conducted extensive ablation studies and interpretability experiments on THU-EP to understand the importance of each module in MACTN and demonstrate its working principles. Finally, we explored the relationship between data quantity and emotion recognition tasks.
\end{itemize}

The organization of the remaining parts of this paper is as follows. Section 2 provides a detailed description of our proposed model. Section 3 describes the dataset and experimental settings. Section 4 presents the results and analysis. We discuss the results in Section 5, and conclude the paper in Section 6.

\section{Proposed Model}
% The conclusion goes here.
% This demo file is intended to serve as a ``starter file''
% for IEEE Computer Society journal papers produced under \Ltransformers
% IEEEtran.cls version 1.8b and later.\cite{mumenthaler2018emotion}
\iffalse
在这一节中，我们具体说明了提出的模型框架，被称为MACTN，如图1(a)所示。为了充分提取情绪的动态性特征，MACTN按层次分成了两个部分，即（1）局部时间特征提取器以及通道注意力部分，（2）全局时间特征提取器部分。在下文中，这些部分将被详细讨论。
\fi
In this section, we specify the proposed model framework, referred to as MACTN, as shown in Figure 1(a). To fully extract the dynamic features of emotions, MACTN is hierarchically divided into two parts, namely (1) the local temporal feature extractor as well as a channel attention part and (2) the global temporal feature extractor part. In the following, these parts will be discussed in detail.

\begin{figure*}[!t]
\centering
\includegraphics[width=0.95\textwidth]{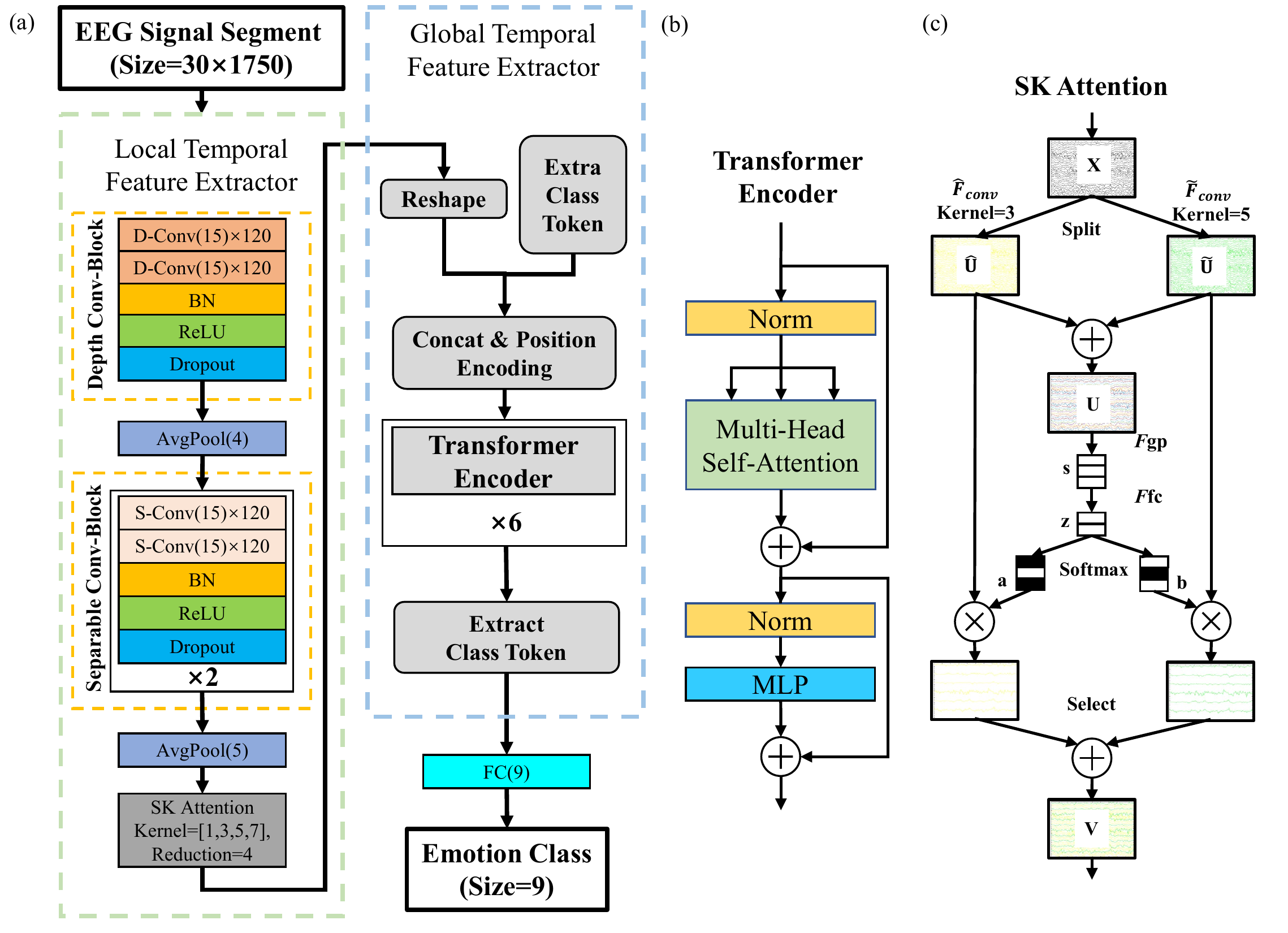}
\caption{Mixed Attention based Convolution and Transformer Network (MACTN) model structure. (a) MACTN consists of 2 main parts, indicated by the green box and the blue box. The green box indicated the low-level local time feature extractor as well as the channel attention part. The blue box indicates the high-level global time feature extractor part. (b) The structure of the Transformer encoder. (c) The structure of Select Kernel (SK) attention (with two streams as an example).}
\label{fig_model}
\end{figure*}

\subsection{Local Temporal Feature Extractor}
\iffalse
Local temporal feature extractor(LTFE)包含3个sub-block，分别为(i)深度卷积块(被称为Depth Conv-Block或D-Conv Block)、(ii)可分离卷积块（被称为Separable Conv-Block或S-Conv Block）和(iii)被称为SK attention的通道注意力计算块[Li, 2019]，SK attention的结构如图1(c)所示。
\fi
Local temporal feature extractor (LTFE) contains three sub-blocks, which are (i)depth convolution block ( denoted as Depth Conv-Block or D-Conv Block), (ii)separable convolution block ( denoted as Separable Conv-Block or S-Conv Block) and (iii)channel attention computation block called SK attention \cite{li2019selective}, and the structure of SK attention is shown in Fig. \ref{fig_model}(c).

\leftline{\textbf{(i) Depth conv-block}}
\iffalse
Depth conv-block主要基于depth-wise temporal convolution 1D(或称D-Conv)，该块主要由5个深度学习层，两个D-Conv层，一个batch norm层，一个ReLU激活函数层，一个dropout层。我们使用深度卷积来为每个输入通道（输入深度）应用一个和/或多个滤波器。
第一个D-Conv层的计算公式如下：
\fi
The Depth conv-block is mainly based on depth-wise temporal convolution 1D (or D-Conv), which consists of five layers, two D-Conv layers, a batch norm (BN) layer, a ReLU activation function layer, and a dropout layer. We use depth-wise convolution to apply one and more filters for each input channel (input depth).
The first D-Conv layer is calculated as follows:

\begin{equation}
\label{eq1}
\begin{array}{c}
H_{k_{2}, t}^{(1)}=W_{k_{1}}^{(1)} X_{m, t: t+P-1}; \\
m=1,2, \ldots, M ; t=1,2, \ldots, T ;  \\
k_{1}=1,2, \ldots, C_{1} ; 
k_{2}=1,2, \ldots, C_{1} M
\end{array}
\end{equation}

\iffalse
 表示经过卷积操作得到的特征， ,  代表滤波器的权重， 。滤波器的长度为 ，并且滤波器的数量为 ，其中 ， 代表输入的原始EEG信号， ， 代表EEG信号的通道数。第一个D-Conv层的计算可以简化为 。第二个D-Conv卷积与第一个D-Conv卷积类似，其计算公式如下：
\fi

\noindent $H$ denotes the feature obtained by convolution operation, $H^{(1)}\in\mathbb{R}^{K_{2}\times T_{1}}$, $W^{(1)}$ denotes the weight of the filter, $W^{(1)}\in\mathbb{R}^{K_{2}\times P}$. The length of the filter is $P$, the number of filters is $K_{2}$, $K_{2}=C_{1}\times M$, $X$  represents the input raw EEG segment, $X\in\mathbb{R}^{M\times T}$, $M$  represents the number of channels of EEG signal. The calculation of the first D-Conv layer can be reduced to $H=\mathrm{DConv_{1}}(X)$. The second D-Convolution is similar to the first D-Convolution and is formulated as follows:
\begin{equation}
% \label{eq2}
\begin{array}{c}
H_{k_{2}, t}^{(2)}=W_{k_{2}}^{(2)} H_{k_{2}, t:t+P-1}^{(1)}; \\
t=1,2, \ldots, T ; k_{2}=1,2, \ldots, K_{2}
\end{array}
\end{equation}

\iffalse
与第一个D-Conv层相比，其输入特征图数量与输出特征图数量相同，而非扩大了C1倍。第二个D-Conv层的计算可以简化为 。
经过两个D-Conv层之后，特征图被施加了非线性的激活层、归一化层以及dropout层操作，这些计算可以使用 表示，因此，depth conv-block的计算可以用以下公式来表达：
\fi
Compared with the first D-Conv layer, the number of input feature maps is the same as the number of output feature maps instead of being extended by a factor of $C_{1}$. The calculation of the second D-Conv layer can be simplified as $H^{(2)}=\mathrm{DConv_{2}}(H^{(1)})$.
After two D-Conv layers, the feature maps are applied with nonlinear activation layer, BN layer, and dropout layer operations, and these calculations can be expressed using $\delta (\bullet )$. Thus the calculation of the depth conv-block can be expressed by the following equation:
\begin{equation}
% \label{eq2}
H_{DConv}=\delta\left(\mathrm{DConv}_{2}\left(\mathrm{DConv}_{1}(X)\right)\right)
\end{equation}

\leftline{\textbf{(ii) Separable conv-block}}
\iffalse
Separable conv-block块主要基于separable temporal convolution 1D(或称S-Conv)，该块主要包含5个层，两个S-Conv层，一个BN层，一个ReLU激活函数层，一个dropout层。整个separable conv-block会重复N次，分别称之为S-Conv block  ， 。
S-Conv层是一个组合的层，每个S-Conv层包含两个卷积，第一个卷积为D-Conv，第二个卷积为pointwise卷积，pointwise卷积的计算公式如下：
\fi
The separable conv-block is mainly based on separable temporal convolution 1D (or S-Conv), which contains five layers, two S-Conv layers, a BN layer, a ReLU layer, and a dropout layer. The whole separable conv-block will be repeated N times, called S-Conv block $S_{i}, S_{i}=1,2,..., N$.
The S-Conv layer is a combined layer, and each S-Conv layer contains two convolutions, the first convolution is D-Conv and the second convolution is a pointwise convolution. The pointwise convolution is calculated as follows:
\begin{equation}
% \label{eq2}
\begin{array}{c}
H_{k_{2}, t}^{(3)}=W_{k_{2}}^{(3)} H_{k_{2}, t}^{(3)};\\
t=1,2, \ldots, T ; k_{2}=1,2, \ldots, K_{2}
\end{array}
\end{equation}

\iffalse
第一个S-Conv层的计算可以简化为 。与depth conv-block类似，每两个S-Conv层之后会添加激活层、归一化层和dropout层，因此第 个S-Conv block的计算可以用以下公式来表达：
\fi
The computation of the first S-Conv layer can be simplified as $H^{(3)}=\mathrm{SConv}_{1}(H^{(2)})$. Similar to the depth conv-block, BN, ReLU, and dropout layers are added after every two S-Conv layers, so the computation of the S-Conv block $S_{i}$  can be formulated as:
\begin{equation}
% \label{eq2}
H_{SConv}^{S_{i}}=\delta\left(\mathrm{SConv}_{2}^{S_{i}}\left(\mathrm{SConv}_{1}^{S_{i}}\left(H_{SCon v}^{S_{i-1}}\right)\right)\right)
\end{equation}

\leftline{\textbf{(iii) Channel attention}}
\iffalse
我们的模型使用了SK attention，channel attention子块也可称之为SK attention子块。与depth conv-block和separable conv-block相比，SK attention子块使用若干核尺寸较小的卷积，并且使用attention机制选择出经过不同卷积计算出的特征图，将它们进行融合。具体来说，我们通过三个操作符——拆分、融合和选择来实现SK卷积，如图1(c)所示，其中显示了两个分支的情况。
\fi
Our model uses SK attention, and the channel attention sub-block can also be called the SK attention sub-block. Compared with the depth conv-block and separable conv-block, the SK attention sub-block uses several convolutions with smaller kernel sizes and uses the attention mechanism to select the feature maps computed by different convolutions and fuse them. Specifically, we implement the SK convolution by three operators – (A) Split, (B) Fuse, and (C) Select, as illustrated in Fig. \ref{fig_model} (c), taking two streams as an example.

\iffalse
Split:给定输入 ，  ，  为经过separable conv-block与平均池化层计算的特征图，  为特征图的特征或者通道数量，  为时间长度。使用两个转换，  和 ，得到 、 ，值得注意的是， 包括卷积、BN和ReLU层。
\fi
(A) Split: Given the input $x$, $x\in\mathbb{R}^{K_{2}\times T' } $, $x$  is the feature map computed after separable conv-block with average pooling layer, $K_{2}$  represents the number of features or the number of channels in the feature map, $ T'$  represents the length of time. Use two transforms, $\mathrm{\hat{F}_{conv}}:x\longrightarrow \hat{U}\in\mathbb{R}^{K_{2}\times T'}$ and $\mathrm{\tilde{F}_{conv}}:x\longrightarrow \tilde{U}\in\mathbb{R}^{K_{2}\times T'}$ , to obtain $\hat{U}$  and $\tilde{U}$ , it is noted that $\mathrm{F_{conv}}$  includes convolutional, BN and ReLU layers.

\iffalse
Fuse: Fuse的基本思想是使用门来控制信息从多个分支流到下一层的神经元，这些分支携带不同规模的信息。为了实现这个目标，首先需要整合多分支信息，通过逐元素求和实现：
\fi

(B) Fuse: The basic idea of Fuse is to use gates to control the flow of information from multiple streams to the next layer of neurons, which carry information of different scales. To achieve this goal, it is first necessary to integrate multi-stream information, which is achieved by element-by-element summation of:
\begin{equation}
% \label{eq2}
U=\widehat{U}+\widetilde{U}
\end{equation}

We then embed the global information by simply averaging over the temporal dimensions to generate the channel statistics as follows:
\begin{equation}
% \label{eq2}
S=\mathrm{F_{gp}}(U)=\frac{1}{T'} \sum_{i=1}^{T'} U_{(i)}
\end{equation}

\iffalse
其中， ，然后通过全连接层实现信息的压缩，该操作记为 
\fi
\noindent where $S \in \mathbb{R}^{K_{2}}$, and then compress the information through the fully connected layer, the operation is noted as $F_{fc}$ :
\begin{equation}
\label{eq_zfc}
z=\mathrm{F_{fc}}(S)=WS
\end{equation}

\iffalse
其中，  ， 是信息压缩权重， ，为了研究 对模型效率的影响，我们使用缩减比 来控制其值， 为固定的数值：
\fi
\noindent in the equation (\ref{eq_zfc}), $z \in \mathbb{R}^{d}$, $W$ is the weight of information compression and $ W \in \mathbb{R}^{d \times K_{2}}$, to control the value of $d$  in order to study its effect on the model efficiency, we use the scaling ratio $r$. $L$ is a constant value.
\begin{equation}
% \label{eq2}
d=\max (K2/r, L)
\end{equation}

\iffalse
Select: 一个跨通道的软注意力被用来自适应地选择不同尺度的信息，这是由紧凑的特征描述符 引导的。 具体来说，使用全连接层和softmax算子提取紧凑通道特征 上的通道信息。
\fi
(C) Select: Cross-channel soft attention is used to adaptively select information at different scales, which is guided by the compact feature descriptor $z$. To be specific, the channel information on the compact channel feature $z$ is extracted using the fully connected layer and the softmax operator.
\begin{equation}
\label{eq_chanattn}
 a=\frac{e^{W_{a} z}}{e^{W_{a} z}+e^{W_{b} z}}, b=\frac{e^{W_{b} z}}{e^{W_{a^{z}}}+e^{W_{b} z}}
\end{equation}

\iffalse
 、 代表 和 的注意力向量。 、 分别代表select权重， 、 。最终的特征图 是通过对不同核的注意权重向量得到的：
\fi
\noindent $a$, $b$  represent the attention vectors of $\hat{U}$ and $\tilde{U}$. $W_{a}$, $W_{b}$ represent the select weights, $W_{a} \in \mathbb{R}^{K_{2} \times d} , W_{b} \in \mathbb{R}^{K_{2} \times d}$ , respectively. The final feature map $v$  is obtained by vectors of attention weights for different kernels:
\begin{equation}
% \label{eq2}
 v=a \widehat{U}+b \widetilde{U}
\end{equation}

\begin{table*}[!t]
    \renewcommand{\arraystretch}{1.3}
    \setlength{\abovecaptionskip}{0cm}
    \caption{Structure of the proposed model}
    \label{table_model}
    \centering
    \begin{threeparttable}
    \begin{tabular}{ccccccc}
    \Xhline{1pt}
    \textbf{Block} & \textbf{Sub-block} & \textbf{Step No.} & \textbf{Step Name} & \textbf{Parameters} & \makecell[c]{\textbf{THU-EP} \\ \textbf{Output Shape}}  & \makecell[c]{\textbf{DEAP} \\ \textbf{Output Shape}} \\ \Xhline{0.4pt}
    \multirow{16}{*}{\makecell[c]{Local\\Temporal\\Feature\\Extractor}} &  & 1 & Input & -- & (30, 1750) & (28, 1536) \\
     & \multirow{2}{*}{\makecell[c]{Depth\\Conv-Block}} & 2 & \makecell[c]{Depth-wise\\Conv1D} & Size=15, Group=Chans, Num=Chans$\times$4 & (120, 1736) & (112, 1522) \\
     &  & 3 & \makecell[c]{Depth-wise\\Conv1D} & \makecell[c]{Size=15, Group=Chans$\times$4, Num=Chans$\times$4,\\ BN, ReLU, Dropout=0.5} & (120, 1722) & (112, 1508) \\
     &  & 4 & AvgPool1D & Pool size=4 & (120, 430) & (112, 377) \\
     & \multirow{4}{*}{\makecell[c]{Separable\\Conv-Block}} & 5 & \makecell[c]{Separable\\ Conv1D} & \makecell[c]{Size=15, Group=Chans$\times$4, Num=Chans$\times$4, \\ Padding=7} & (120, 430) & (112, 377) \\
     &  & 6 & \makecell[c]{Separable\\Conv1D} & \makecell[c]{Size=15, Group=Chans$\times$4, Num=Chans$\times$4, \\ Padding=7, BN, ReLU, Dropout=0.5} & (120, 430) & (112, 377) \\
     &  & 7 & \makecell[c]{Separable\\Conv1D} & \makecell[c]{Size=15, Group=Chans$\times$4, Num=Chans$\times$4, \\ Padding=7} & (120, 430) & (112, 377) \\
     &  & 8 & \makecell[c]{Separable\\Conv1D} & \makecell[c]{Size=15, Group=Chans$\times$4, Num=Chans$\times$4, \\ Padding=7, BN, ReLU, Dropout=0.5} & (120, 430) & (112, 377) \\
     &  & 9 & AvgPool1D & Pool size=5 & (120, 86) & (112, 75) \\
     & \makecell[c]{Channel\\Attention} & 10 & SK Attention & Kernel size={[}1,3,5,7{]}, Reduction=4 & (120, 86) & (112, 75) \\ \Xhline{0.4pt}
    \multirow{9}{*}{\makecell[c]{Global\\Temporal\\Feature\\Extractor}} &  & 11 & Reshape & -- & (86, 120) & (75, 112) \\
     &  & 12 & Extra Class Token & Input shape=(1, Chans$\times$4) & (1, 120) & (1, 112) \\
     &  & 13 & \makecell[c]{Concatenate \& \\Position Encoding} & -- & (87, 120) & (76, 112)\\
     & Transformer & 14-19 & \makecell[c]{Temporal\\Transformer\\Encoder} & \makecell[c]{Head=8, Head dim=256, \\ MLP dim=128} & (87, 120) & (75, 112)\\
     &  & 20 & Extract Class Token & -- & 120 & 112 \\
     &  & 21 & FC & -- & 9 & 2 \\ \Xhline{1pt}
    \end{tabular}
        \begin{tablenotes}
        % \footnotesize
        \item \textit{Chans: The number of channels in THU-EP is 30, while in DEAP it is 28.}
        \end{tablenotes}
        \end{threeparttable}
    \end{table*}

\subsection{Global Temporal Feature Extractor}
\iffalse
GTFE由若干个transformer encoder组成，图1(b)为transformer encoder的结构，除此之外，GTFE还包括添加可学习的class token、位置编码、提取class token等操作。
Transformer encoder主要包括两个模块，多头自注意力（MHSA）和前馈网络（MLP），在每个模块之前，会应用Layer Normalization（LN），残差连接会应用在每个模块之后。MHSA可以被描述为使用三个matrix，query（Q）、key（K）和value（V）之间计算Scaled Dot-Product Attention，我们计算输出为：
\fi
Global temporal feature extractor (GTFE) consists of several transformer encoders, Fig. \ref{fig_model}(b) shows the structure of the transformer encoder. Besides, GTFE includes operations such as adding a learnable class token, position encoding, and extracting the class token.
The transformer encoder mainly consists of two modules, multi-headed self-attentive (MHSA) and feedforward network (MLP), where layer normalization (LN) is applied before each module, and residual connectivity is applied after each module. MHSA can be described as using three matrices, query (Q), key (K), and value (V), to calculate scaled dot-product attention among them, and we calculate the output as:

\begin{equation}
\label{eq_sa}
  A=\mathrm{Attention}(Q, K, V)=\mathrm{softmax}\left(\frac{Q K^{T}}{\sqrt{d_{k}}}\right) V
\end{equation}

\iffalse
其中 表示尺寸放缩因子， 表示query和/或key的维度。用不同的学习线性投影将查询、键和值线性投影 次到 、 和 维度是有益的。多头自注意力可以被表示为：
\fi

\noindent where $1/\sqrt{d_{k}}$  denotes the dimensional deflation factor and $d_{k}$  denotes the dimensionality of query and/or key. It is beneficial to linearly project the query, key, and value $h$  times to the $d_{k}$, $d_{k}$, and $\ d_{v}$ dimensions using different learned linear projections. Multi-head self-attention can be represented as:
\begin{equation}
% \label{eq2}
\begin{array}{c}
\mathrm{MHSA}(x)=\mathrm{concat}\left(A_{1}, A_{2}, \ldots, A_{h}\right) W^{O}, \\
\text { where } A_{i}=\mathrm { Attention }\left(x W_{i}^{Q}, x W_{i}^{K}, x W_{i}^{V}\right)
\end{array}
\end{equation}

\iffalse
其中投射器是参数矩阵 ， ， 。
前馈网络MLP的计算可以用如下公式表示：
\fi
\noindent the projectors are the parameter matrices $W^{Q}_{i} \in \mathbb{R}^{(d_{seq}+1) \times d_{embed} \times d_{k}}$, $W^{O} \in \mathbb{R}^{(d_{seq}+1) \times hd_{v} \times d_{embed}}$, $d_{k}=d_{v}$.
The calculation of the feedforward network MLP can be expressed as follows:
\begin{equation}
% \label{eq2}
  \mathrm{MLP}(x)=\delta\left(x W_{1}^{MLP}+b_{1}^{M L P}\right) W_{2}^{M L P}+b_{2}^{M L P}
\end{equation}

\iffalse
其中前馈网络的转换参数矩阵 ， 。整个transformer encoder的计算输出为：
\fi
\noindent Where the transformation parameter matrix of the feedforward network $W^{MLP}_{1} \in \mathbb{R}^{(d_{seq}+1) \times d_{embed} \times d_{MLP}}$ , $W^{MLP}_{2} \in \mathbb{R}^{(d_{seq}+1) \times d_{MLP} \times d_{embed}}$ . The computed output of the entire transformer encoder is:

\begin{equation}
% \label{eq2}
\begin{array}{c}
T E^{i}=\mathrm{MLP}\left(\mathrm{LN}\left(S A^{i}\right)\right)+S A^{i} \\
\text { where } S A^{i}=\mathrm{MHSA}\left(\mathrm{LN}\left(T E^{i-1}\right)\right)
\end{array}
\end{equation}

\iffalse
其中第 层的transformer encoder输出为 ，自注意力输出为 。在论文中，在送入transformer encoder之前，对序列逐元素添加一个可学习的编码作为位置编码，就像BERT中那样[Devlin，2018]。
\fi
\noindent where the transformer encoder output of layer $i$  is $TE^{i}$  and the self-attentive output is $SA^{i}$. In the paper, a learnable encoding is added to the sequence element by element as a position encoding before feeding into the transformer encoder, as in BERT \cite{devlin2018bert}.

\FloatBarrier
\section{Experiment}
\subsection{Dataset}
\iffalse
para 1 exp
\fi
In order to evaluate MACTN, we conducted several experiments on two public datasets, one being the Tsinghua University Emotional Profiles (THU-EP) and the other being the Database for Emotion Analysis using Physiological signals (DEAP) \cite{koelstra2011deap}.

\iffalse
para 2 exp
\fi
The THU-EP database includes 80 college students (50 females and 30 males) with an average age of 20.16 years old ranging from 17 to 24 years old. There are 28 video clips used as stimuli, which include 9 different emotions: anger, disgust, fear, sadness, amusement, joy, inspiration, tenderness, and neutral. Except for the 4 video clips used to induce neutral emotions, each of the other emotions was induced using 3 video clips. Each subject viewed the video clips in seven blocks, with each block containing four trials. Participants were asked to solve 20 arithmetic problems between two blocks\cite{li2020eeg}. NeuSen.W32 wireless EEG system was used to record EEG signals with a sampling frequency of 250Hz. The positions and names of the 32 channels are shown in Fig. \ref{fig_preprocess}. 

\iffalse
para 3 exp
\fi
DEAP is a human emotional dataset that contains multiple physiological signals, including EEG and galvanic skin response (GSR). The dataset consists of 32 participants (17 males and 15 females) with an average age of 27.19 years ranging from 19 to 37 years old. 40 music videos, each lasting one minute, were used as stimuli. Each subject completed 40 trials, and after each trial, they were asked to rate their emotional state using four dimensions: arousal, valence, dominance, and liking, each rated on a discrete scale ranging from 1 to 9. The EEG was recorded using a 32-channel device at a sampling frequency of 512 Hz.

\subsection{Pre-process}
\begin{figure*}[!t]
\centering
\includegraphics[width=0.7\textwidth]{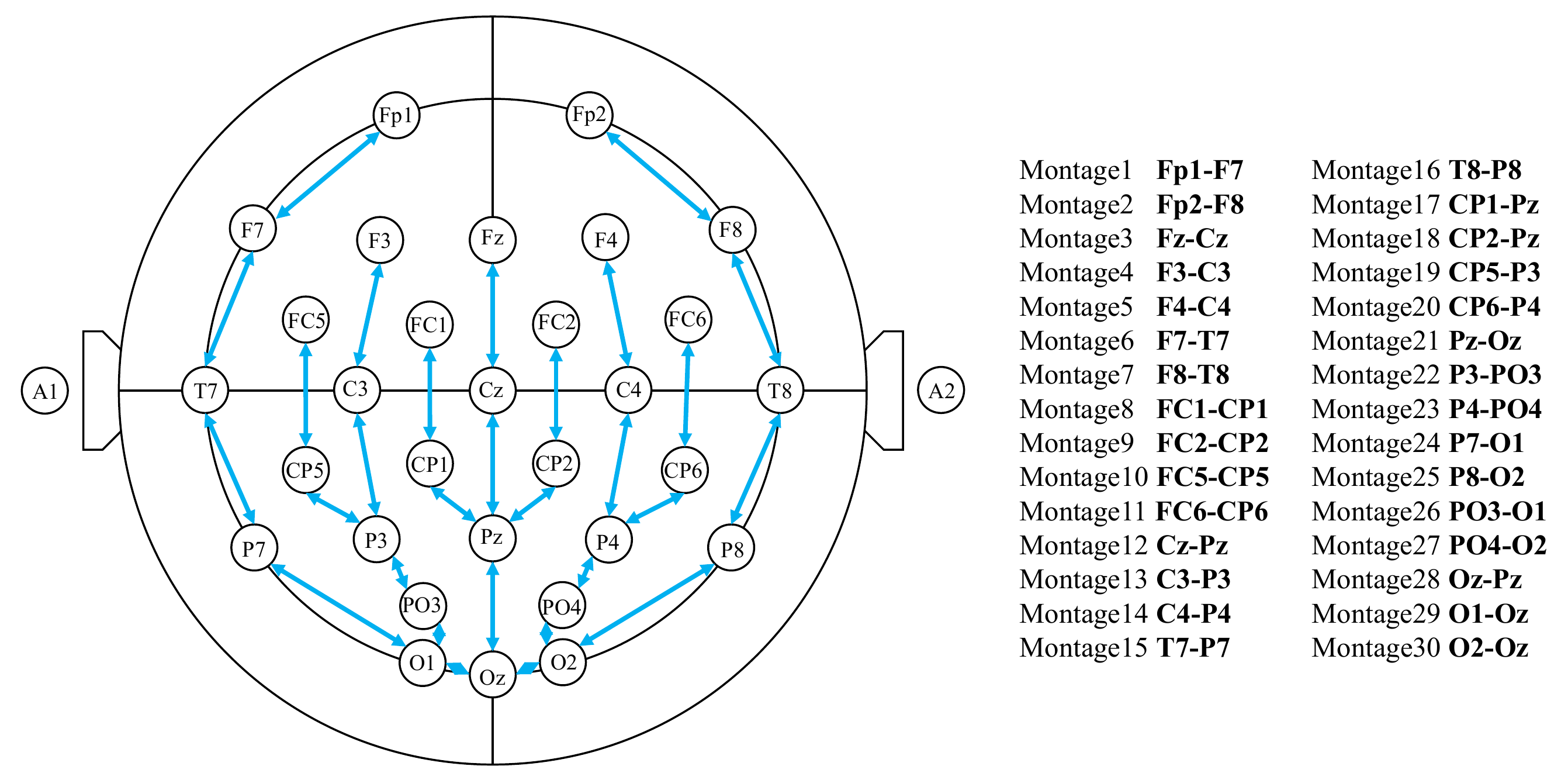}
\caption{Bipolar re-referencing montage.}
\label{fig_preprocess}
\end{figure*}

\iffalse
para 4 exp
\fi
 The same preprocessing methods were applied to each trial for THU-EP. Firstly, a bipolar re-referencing method was used as shown in Fig. \ref{fig_preprocess}. The electrodes from the left and right mastoids were discarded (as the left and right mastoid signals were not recorded in the BCI competition), and the remaining 30 channels were reconfigured into 30 montages by subtracting each channel pairwise. Montage 21 and Montage 28 contained signals that were symmetrical about 0 uV. Secondly, each EEG segment was extracted using a sliding window approach with a window length of 14 seconds and a step size of 4 seconds. Thirdly, a 6th order Butterworth filter was used for filtering, which included a 50 Hz notch filter (with a notch width of 48-52 Hz) and a bandpass filter of 0.5-45 Hz. Fourthly, to simplify the subsequent calculations, the EEG signals were downsampled to 125 Hz. Fifthly, a Z-score normalization method was used to normalize each segment. After preprocessing, each subject had a total of 363 EEG segments (min=362, max=366, mode=363).
\FloatBarrier

\iffalse
para 5 exp
\fi
For DEAP, the EEG data from each trial were downsampled to 128Hz, eye artifacts were removed, and bandpass filtering was performed using a 4-45Hz filter. The filtered signals were then re-referenced using common average referencing. The DEAP dataset provides signals that have been processed according to the above steps. We partitioned each trial into 13 segments using a sliding window approach with a window length of 12 seconds and a step size of 4 seconds. Following previous work, we converted the ratings for arousal and valence dimensions into binary labels using a threshold of 5. We extracted 28 channels based on the TSception setup and rearranged them accordingly \cite{ding2022tsception}.

\subsection{Implement Details}
\iffalse
para 6 exp
\fi
In the LTFE of the model, all operations are 1-dimensional (e.g., convolution, BN, pooling, etc.), the size of the convolution kernels are all 15, the number is fixed at 120 (i.e., $K_2$ = 120 in THU-EP or $K_2$ = 120 in DEAP) in THU-EP and 112 in DEAP. in Depth ConvBlock, $C_1$ is set to 4, i.e., in Step No. 2 of Table \ref{table_model}, four separate convolutions are used to process the data for each EEG channel. Depth ConvBlock does not use padding for all convolutions, and after two convolutions, the output feature map is reduced by 28 in the temporal dimension. After Depth ConvBlock, we use an average pooling layer to compress the temporal features (i.e., Stem No. 4) with a pooling size of 4. Unlike Depth ConvBlock, Separable Convolution in ConvBlock uses padding to ensure that the temporal dimension is invariant before and after the computation. Similarly, after Separable ConvBlock, we compress the features using an average pooling layer with a pooling size of 5. After the average pooling, SK attention was used in which we used four convolutions with smaller kernel sizes, 1, 3, 5, and 7, to extract finer-grained temporal features and integrate them, with the compression ratio $r$ set to 4 and $L$ set to 32 in keeping with SKNet\cite{li2019selective}.

\iffalse
para 7 exp
\fi
In GTFE, the learnable class token and position embedding are initialized using standard Gaussian distribution. We use six layers of transformer encoder, and in each layer of transformer encoder, we set the number of multi-head self-attention heads to 8 and set the dimensions of Q, K, and V to $d_{k}=d_{v}=256$. It is noted that $d_k$ is not equal to $d_{embed}/h$, and $d_{embed}$ is 120 (THU-EP) or 112 (DEAP), while $d_{seq}$ is 86 (THU-EP) or 75 (DEAP). The dimension of the internal layer of the feedforward network is set to 128.

\iffalse
para 8 exp
\fi
In the experiments, the optimizer applied is AdamW optimizer \cite{loshchilov2018decoupled}, the initial learning rate is set to 0.001, the learning rate decay strategy is ReduceOnPlateau, i.e., the loss decays to 10\% of the original after ten epochs without decreasing, the weight decay is set to 0.0001, the batch size is 16, the number of epochs is 100, and the early stopping strategy is applied. At the same time, we use the flooding strategy \cite{ishida2020we} to prevent the model from overfitting and set its parameter b to 1.3. Our model is implemented in Python 3.8 using PyTorch 1.10. The model is configured to run on a GeForce RTX 3090 GPU. The entire model code can be accessed on GitHub\footnote{https://github.com/ThreePoundUniverse/MACTN}.

\subsection{Performance Evaluation}
\iffalse
para 9 exp
\fi
For THU-EP, we employed three different cross-validation methods: 10-fold cross-subject-validation (10F-CSV), leave-one-subject-out cross-validation (LOSO), and leave-one-trial-out cross-validation (LOTO). To facilitate comparison, the data partitioning and performance evaluation of 10F-CSV were kept consistent with Shen et al. \cite{shen2022contrastive}, where some subject data were used for training and the remaining data were used for testing. Each subject in THU-EP participated in 28 trials, and LOTO involved selecting one trial for testing while using the other 27 trials for training. This process was repeated 28 times to ensure that all 28 trials were tested. To investigate the impact of the number of subjects on the performance of MACTN, we added the LOSO method. Unlike the 10F-CSV method, we divided all subjects into three sets: train, validation, and test. One subject was used for testing, and the data of the remaining subjects were randomly divided into 80\% and 20\%. For example, with 80 subjects, the number of subjects in the training set, validation set, and test set were 63, 16, and 1 (80 = 63 + 16 + 1). This process was repeated 80 times.

\iffalse
para 10 exp
\fi
For DEAP, we adopted two methods for cross-validation: 10-fold cross-trial-validation (10F-CTV) and leave-one-subject-out cross-validation (LOSO). To facilitate comparison, the data partition and performance evaluation of 10F-CTV were consistent with Ding et al. \cite{ding2022tsception}, with 40 experiments, in which 4 were used for testing and the other 36 were split into training and validation sets based on the proportion of 80\% and 20\% of the segments, repeated 10 times. The data partition of LOSO was consistent with Zhang et al. \cite{zhang2022unsupervised}, with one subject used for testing and the remaining subjects used for training, repeated 32 times.

The evaluation metrics for model performance are accuracy (ACC) and F1 score (F1).
% \begin{equation}
% % \label{eq2}
% \begin{array}{c}
% macroP = \frac{1}{n}\sum\limits_{i = 1}^n {{P_i}} ,
% {\text{where }}{P_i} = \frac{{T{P_i}}}{{T{P_i} + F{P_i}}} \\
% macroR=\frac{1}{n}\sum\limits_{i = 1}^n {{R_i}} , \text {where } R_{i}=\frac{T P_{i}}{T P_{i}+F N_{i}}
% \end{array}
% \end{equation}
% \begin{equation}
% % \label{eq2}
% F1 = \frac{{2 \times macroP \times macroR}}{{macroP + macroR}}
% \end{equation}

% \iffalse
% 我们使用 、 、 、 表示每一类的假阳性、真阳性、假阴性和真阴性。
% \fi
% $FP_{i}$, $TP_{i}$, $FN_{i}$, and $TN_{i}$ denote false positives, true positives, false negatives, and true negatives for each class.

\subsection{MACTN Feature Explainability}
\iffalse
para 10-15 exp
\fi
In this paper, we use four methods to explain our model based on the THU-EP dataset. These methods are as follows: 1) feature map visualization, 2) convolution kernel visualization, 3) channel attention weight visualization, and 4) self-attention weight visualization, where the fourth method is dedicated to explaining the interpretability of individual segments.

1) Feature map visualization: We use T-SNE to reduce the dimensions of specific feature maps and create a scatter plot to display the topological relationships between each feature map.

2) Convolution kernel visualization: This method focuses on directly visualizing and interpreting the convolution kernel weights from the model. In our proposed model, we use temporal convolutions, and due to the limited connectivity of the convolution layers (using depthwise and separable convolutions), temporal convolutions can be interpreted as narrowband frequency filters. Additionally, smaller kernels between adjacent convolution layers can be equivalently represented as larger convolutions using a calculation formula to extract richer frequency information.

3) Channel attention weight visualization: This method focuses on directly visualizing and interpreting the attention weights of the channel dimension to display the model's preference for specific channels. Similarly, due to the limited connectivity of the convolution layers in our proposed model, each original EEG channel corresponds to several ($C_{1}$) feature channels after convolution calculation. Therefore, averaging every $C_{1}$ feature channel can demonstrate the model's preference for specific EEG channels and corresponding brain regions through channel attention visualization.

4) Self-attention weight visualization: This method focuses on directly visualizing and interpreting the attention weights of the temporal dimension. In our proposed model, we use a temporal transformer that can focus on different emotional information on a global temporal scale and assign higher attention weights. Notably, these high-weight assignments are sparse, allowing the model to provide certain interpretability when combined with the original stimuli.

\FloatBarrier
\section{Results}
\iffalse
para 1 res
\fi
In this section, we report and compare our results in terms of accuracy and F1 score with state-of-the-art methods. We then conduct an ablation study to reveal the role of each component in MACTN.

\subsection{Emotion Recognition Performance}
\iffalse
para 2 res
\fi
We compared our proposed method with the state-of-the-art (SOTA) methods that also applied the THU-EP dataset. Table \ref{table_thuep_subj} shows the 10-fold cross-subject-validation results on THU-EP, where our method achieved the best performance with an 11.6\% higher accuracy than the CLISA method proposed by Shen et al.\cite{shen2022contrastive}.  Moreover, we compared our method with other convolutional neural network-based models, including DeepConvNet \cite{lawhern2018eegnet}, ShallowConvNet \cite{lawhern2018eegnet}, EEGNet \cite{lawhern2018eegnet}, and TSception \cite{ding2022tsception}, in terms of accuracy and F1 score. Our proposed method outperformed these models by 20\% in the non-transfer learning setting, and it is worth noting that some parameter settings of these models are consistent with their proposers. Among the many results, we observed that the performance of these models was lower than that of CLISA, whose model complexity was similar to other models. We found that increasing the model complexity of DeepConvNet could still improve its performance. Table \ref{table_thuep_trl} shows the leave-one-trial-out cross-validation results on THU-EP, where our proposed MACTN method achieved 62.5\% accuracy and 61.6\% F1 score. Compared to DeepConvNet, ShallowConvNet, EEGNet, TSception, our method achieved an 8\% accuracy improvement. Compared to SVM and KNN, MACTN surpassed their accuracy by more than 20\%.

\begin{table}[!t]
    \renewcommand{\arraystretch}{1.3}
    \setlength{\abovecaptionskip}{0cm}
    \caption{The accuracies (ACC) and F1 scores(mean$\pm$STD) on the THU-EP dataset  (cross-subject)}
    \label{table_thuep_subj}
    \centering
    \begin{threeparttable}
    \setlength{\tabcolsep}{16pt}{
    \begin{tabular}{ccc}
    \Xhline{1pt}
    \textbf{Model} & \textbf{ACC (\%)} & \textbf{F1 (\%)} \\ \Xhline{0.4pt}
    KNN & 25.1 $\pm$ 4.5 & 25.0 $\pm$ 4.5 \\
    SVM & 27.5 $\pm$ 3.1 & 27.3 $\pm$ 3.2 \\ \Xhline{0.4pt}
    DeepConvNet & 35.6 $\pm$ 4.8 & 34.4 $\pm$ 5.1 \\
    ShallowConvNet & 31.2 $\pm$ 3.1 & 30.8 $\pm$ 3.2 \\
    EEGNet & 33.4 $\pm$ 3.8 & 32.0 $\pm$ 4.1 \\
    TSception & 35.8 $\pm$ 4.1 & 35.2 $\pm$ 4.3 \\
    CLISA \cite{shen2022contrastive} & 47.0 $\pm$ 4.4 & - \\ \Xhline{0.4pt}
    MACTN (Proposed) & \textbf{58.6 $\pm$ 5.7} & \textbf{58.7 $\pm$ 5.9} \\ \Xhline{1pt}
    \end{tabular}
    }
    
        % \begin{tablenotes}
        % \item[$\sharp$] \textit{The result is sourced from the presented paper.}
        % \end{tablenotes}
        \end{threeparttable}
    \end{table}

\begin{table}[!t]
    \renewcommand{\arraystretch}{1.3}
    \setlength{\abovecaptionskip}{0cm}
    \caption{The accuracies (ACC) and F1 scores(mean$\pm$STD) on the THU-EP dataset (cross-trial)}
    \label{table_thuep_trl}
    \centering
    \begin{threeparttable}
    \setlength{\tabcolsep}{16pt}{
    \begin{tabular}{ccc}
    \Xhline{1pt}
    \textbf{Model} & \textbf{ACC (\%)} & \textbf{F1 (\%)} \\ \Xhline{0.4pt}
    KNN & 40.2 $\pm$ 7.8 & 40.1 $\pm$ 7.9 \\
    SVM & 40.1 $\pm$ 7.6 & 39.9 $\pm$ 7.7 \\ \Xhline{0.4pt}
    DeepConvNet & 49.9 $\pm$ 7.1 & 49.4 $\pm$ 7.2 \\
    ShallowConvNet & 44.0 $\pm$ 5.9 & 43.1 $\pm$ 5.4 \\
    EEGNet & 49.6 $\pm$ 6.8 & 49.1 $\pm$ 6.9 \\
    TSception & 54.3 $\pm$ 8.0 & 53.1 $\pm$ 7.7 \\ \Xhline{0.4pt}
    MACTN (Proposed) & \textbf{62.5 $\pm$ 7.1} & \textbf{61.6 $\pm$ 7.0} \\ \Xhline{1pt}
    \end{tabular}
    }
        \end{threeparttable}
    \end{table}

\iffalse
para 3 res
\fi
In DEAP, we analyzed based on the F1 score. Compared to accuracy, F1 score is a more reliable metric that can quantify the performance of classification methods when there are imbalanced classes in the dataset. Looking at Table \ref{table_deap_trl}, deep learning methods are generally superior to non-deep learning methods, especially MACTN which outperforms by about 7\% in F1 score. MACTN achieved a arousal F1 score of 65.3\% and a valence F1 score of 65.2\%. TSception method ranks second among other methods with arousal F1 score of 64.1\% and valence F1 score of 64.6\%, which is about 7\% better in F1 score. Among the compared deep learning methods, MACTN leads TSception by 1.2\% in arousal F1 score and 0.6\% in valence F1 score. Compared to other methods such as DeepConvNet, ShallowConvNet, EEGNet, MACTN outperforms by about 3\% in F1 score. The results of the LOSO experiment (Table \ref{table_deap_subj}) also show that deep learning models outperform non-deep learning models in performance. Compared to non-deep learning models, MACTN improves by 6\% in arousal F1 score and 11\% in valence F1 score. Compared to other deep learning models, MACTN improves by 1\% in arousal F1 score and 4\% in valence F1 score.

\begin{table}[!t]
\renewcommand{\arraystretch}{1.3}
\setlength{\abovecaptionskip}{0cm}
\caption{The accuracies (ACC) and F1 scores(mean$\pm$STD) on the DEAP dataset (cross-trial)}
\label{table_deap_trl}
\centering
    \begin{threeparttable}
    \setlength{\tabcolsep}{2pt}{
        \begin{tabular}{ccccc}
            \Xhline{1pt}
            \multirow{2}{*}{\textbf{Model}} & \multicolumn{2}{c}{ \textbf{Arousal}} & \multicolumn{2}{c}{\textbf{Valence}} \\
                & \textbf{ACC (\%)} & \textbf{F1 (\%)} & \textbf{ACC (\%)} & \textbf{F1 (\%)} \\ \Xhline{0.4pt}
            KNN & 60.2$\pm$12.3 & 58.3$\pm$25.0 & 54.3$\pm$9.2 & 56.3$\pm$16.2 \\
            SVM & 61.2$\pm$12.3 & 58.4$\pm$26.6 & 56.3$\pm$7.0 & 58.8$\pm$11.3 \\ \Xhline{0.4pt}
            DeepConvNet & 62.0$\pm$8.6 & 62.6$\pm$17.4 & 60.7$\pm$7.9 & 63.2$\pm$11.6 \\
            ShallowConvNet & 62.1$\pm$10.3 & 62.2$\pm$20.1 & 60.2$\pm$8.3 & 63.1$\pm$10.3 \\
            EEGNet & 59.1$\pm$8.6 & 61.5$\pm$15.2 & 55.7$\pm$8.2 & 58.7$\pm$10.5 \\
            TSception & 63.0$\pm$11.0 & 64.1$\pm$16.6 & 61.6$\pm$8.6 & 64.6$\pm$10.2 \\ \Xhline{0.4pt}
            MACTN (Proposed) & \textbf{63.6$\pm$10.5} & \textbf{65.3$\pm$16.1} & \textbf{61.9$\pm$8.2} & \textbf{65.2$\pm$9.8} \\ \Xhline{1pt}
        \end{tabular}
    }
    \end{threeparttable}
\end{table} 

\begin{table}[!t]
\renewcommand{\arraystretch}{1.3}
\setlength{\abovecaptionskip}{0cm}
\caption{The accuracies (ACC) and F1 scores(mean$\pm$STD) on the DEAP dataset (cross-subject)}
\label{table_deap_subj}
\centering
    \begin{threeparttable}
    \setlength{\tabcolsep}{2pt}{
        \begin{tabular}{ccccc}
            \Xhline{1pt}
            \multirow{2}{*}{\textbf{Model}} & \multicolumn{2}{c}{ \textbf{Arousal}} & \multicolumn{2}{c}{\textbf{Valence}} \\
                & \textbf{ACC (\%)} & \textbf{F1 (\%)} & \textbf{ACC (\%)} & \textbf{F1 (\%)} \\ \Xhline{0.4pt}
                KNN & 59.3$\pm$12.6 & 57.2$\pm$25.0 & 54.1$\pm$8.3 & 50.2$\pm$20.4 \\
                SVM & 60.1$\pm$12.5 & 57.8$\pm$26.6 & 55.6$\pm$7.9 & 51.2$\pm$22.1 \\ \Xhline{0.4pt}
                DeepConvNet & 65.0$\pm$10.0 & 62.5$\pm$23.6 & 56.3$\pm$8.1 & 45.9$\pm$24.5 \\
                ShallowConvNet & 66.2$\pm$13.2 & 60.9$\pm$25.3 & 61.6$\pm$6.7 & 58.5$\pm$16.6 \\
                EEGNet & 60.6$\pm$9.2 & 60.9$\pm$24.5 & 57.1$\pm$7.1 & 51.3$\pm$19.9 \\
                DGCNN \cite{zhang2022unsupervised} & 61.1 & - & 59.3 & - \\
                TSception & 63.1$\pm$8.8 & 60.1$\pm$22.0 & 63.6$\pm$7.8 & 60.5$\pm$21.7 \\ \Xhline{0.4pt}
                MACTN (Proposed) & \textbf{67.8$\pm$8.1} & \textbf{63.4$\pm$22.3} & \textbf{66.1$\pm$6.1} & \textbf{62.2$\pm$18.6} \\ \Xhline{1pt}
        \end{tabular}
    }
    \end{threeparttable}
\end{table} 

% \begin{figure}[!t]
% \centering
% \includegraphics[width=0.5\textwidth]{figures/fig_3_resize.pdf}
% \caption{Proposed model 10-fold cross-validation average confusion matrix.}
% % \label{fig_sim}
% \end{figure}
\iffalse
para 4 res
\fi
According to extensive comparisons with various methods, this approach has demonstrated excellent performance in emotion recognition tasks and exhibits a certain degree of generality.

\subsection{Ablation Study and Parameter Analysis}
\iffalse
para 5 res
\fi
To investigate how our model achieves such performance, we conducted various ablation experiments on the proposed model. Table \ref{table_alab1} examines the contributions of the LTFE and GTFE blocks to the model's performance improvement. The experimental results show that without LTFE, the model only achieved an accuracy of 22\%, while without GTFE, the accuracy decreased by 22.4\% compared to the model with both blocks. This suggests that LTFE plays a more significant role in feature extraction and representation, while the lower performance of the model with only GTFE may be due to the transformer's limited ability to handle longer sequence inputs.

\begin{table}[!t]
\renewcommand{\arraystretch}{1.3}
\setlength{\abovecaptionskip}{0cm}
\caption{For the results of ablation experiments applying local and global feature extractor}
\label{table_alab1}
\begin{tabular}{cccc}
\Xhline{1pt}
\multicolumn{2}{c}{\textbf{Module}} & \multicolumn{2}{c}{\textbf{Metrics}} \\
\begin{tabular}[c]{@{}c@{}}Local Temporal \\ Feature Extractor\end{tabular} & \begin{tabular}[c]{@{}c@{}}Global Temporal \\ Feature Extractor\end{tabular} & ACC (\%) & F1 (\%) \\ \Xhline{0.4pt}
\ding{56} & \ding{52} & 22.0±2.1 & 20.3±2.1 \\
\ding{52} & \ding{56} & 46.2±5.0 & 46.1±5.0 \\
\ding{52} & \ding{52} & 58.6±5.7 & 58.7±5.9 \\ \Xhline{1pt}
\end{tabular}
\end{table}

\iffalse
para 6 res
\fi
Furthermore, we investigated which sub-block in LTFE has a greater impact on the model's performance. In this ablation study, we retained GTFE and compared the models formed by removing any one of the three sub-blocks present in LTFE. The results in Table \ref{table_alab2} indicate that removing the Separable Conv-Block results in the maximum decrease in accuracy, by 16.9\%, compared to when all three sub-blocks are present. Removing the SK attention, on the other hand, only leads to a 2.4\% decrease in accuracy. The model without Depth Conv-Block achieves an accuracy of 46.2\%. It is noteworthy that the Depth Conv-Block is responsible for amplifying the feature channel dimension, and thus removing it results in feature maps with the same channel dimension as the preprocessed EEG channel dimension.

\begin{table}[!t]
\renewcommand{\arraystretch}{1.3}
\setlength{\abovecaptionskip}{0cm}
\caption{For the results of ablation experiments applying the sub-blocks in local temporal feature extractor}
\label{table_alab2}
\begin{tabular}{ccccc}
\Xhline{1pt}
% \IEEEeqnarraymulticol{3}{t}{\textbf{Module}} & \IEEEeqnarraymulticol{2}{t}{\textbf{Metrics}} \\
\multicolumn{3}{c}{\textbf{Module}} & \multicolumn{2}{c}{\textbf{Metrics}} \\
\begin{tabular}[c]{@{}c@{}}Depth \\ Conv-Block\end{tabular} & \begin{tabular}[c]{@{}c@{}}Separable \\ Conv-Block\end{tabular} & \begin{tabular}[c]{@{}c@{}}Channel \\ Attention\end{tabular} & ACC (\%) & F1 (\%) \\ \Xhline{0.4pt}
\ding{56} & \ding{52} & \ding{52} & 46.2±6.5 & 45.9±6.8 \\
\ding{52} & \ding{56} & \ding{52} & 41.7±4.9 & 41.5±4.9 \\
\ding{52} & \ding{52} & \ding{56} & 56.2±5.3 & 56.1±5.4 \\
\ding{52} & \ding{52} & \ding{52} & 58.6±5.7 & 58.7±5.9 \\ \Xhline{1pt}
\end{tabular}
\end{table}
\FloatBarrier

\begin{figure}[htbp]
\centering
\includegraphics[width=0.48\textwidth]{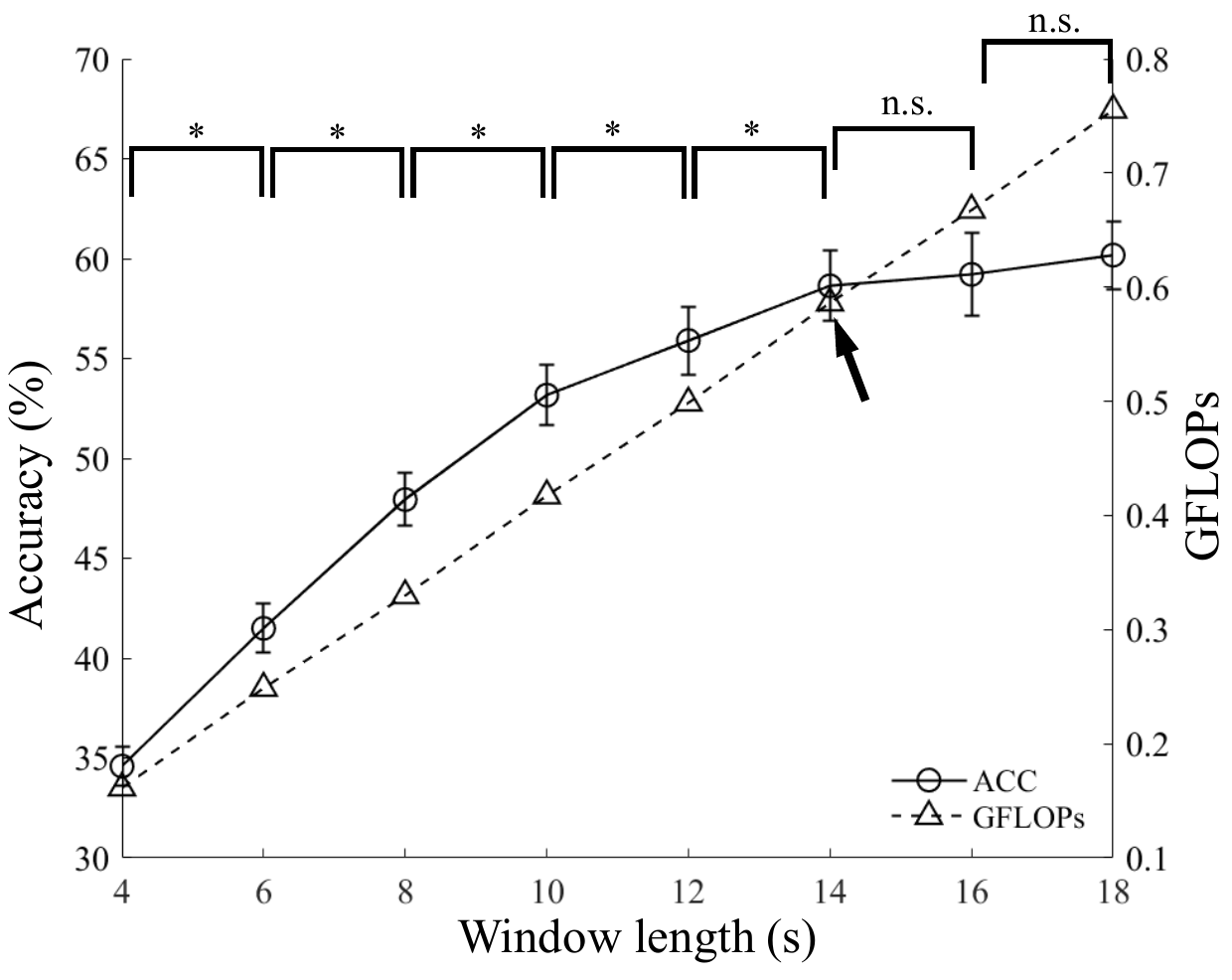}
\caption{Fold plots of MACTN for segments with different window lengths. The error bar indicates the standard deviation of the mean (SEM). Accuracy was tested using the Wilcoxon signed-rank test, with $\ast $ representing $p < 0.05$ and n.s. representing not significant.}
\label{fig_para}
\end{figure}
% \FloatBarrier

\iffalse
para 7 res
\fi
We found that the length of EEG segments can affect the model's accuracy and F1 score. Fig. \ref{fig_para} shows that, with a fixed preprocessing pipeline and sliding window stride of 4 seconds, the model's accuracy consistently increases with the window length. However, the rate of change in accuracy slows down after a certain window length. After conducting a Wilcoxon sign rank test, the model accuracy for input windows of 14 seconds was significantly different from those of 12, 10, 8, 6, and 4 seconds, with p-values less than 0.05. In contrast, the accuracy of the 14-second input was not significantly different from those of 16 and 18 seconds. At the same time, we computed the model's computational complexity (GFLOPs) under different window lengths, and Fig. \ref{fig_para} shows a linear relationship between the window length and computational complexity.

\section{Discussion}
\subsection{Impact of Dataset Size on Classification Performance}
\iffalse
para 5-6 dis
\fi
\begin{figure}[htbp]
\centering
\includegraphics[width=0.45\textwidth]{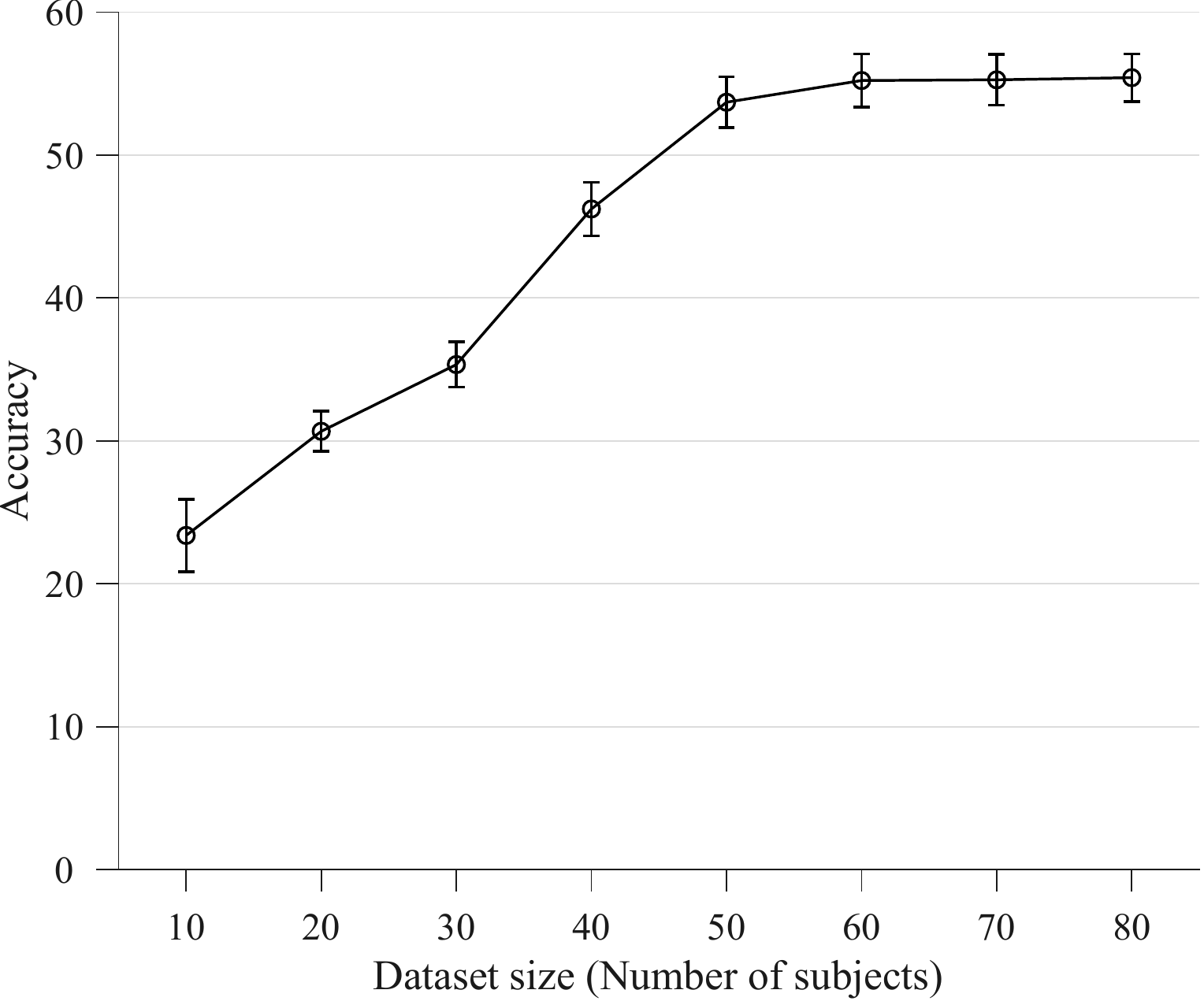}
\caption{ Line chart of average accuracy of MACTN under different dataset sizes.}
\label{fig_num_subj}
\end{figure}
% \FloatBarrier

The THU-EP dataset has a sufficient number of subjects that enables the model to learn individual differences between subjects and narrow them down. We were curious about how the number of subjects affects the performance of MACTN, so we conducted experiments by varying the number of subjects as the independent variable. We selected 10, 20, 30, 40, 50, 60, 70, and all subjects from a pool of 80 subjects and performed leave-one-subject-out cross-validation experiments. The data partitioning and evaluation methods are described in detail in Section 3.

Fig. \ref{fig_num_subj} shows the relationship between the number of subjects and the performance of MACTN. As observed from the figure, the accuracy of MACTN increases with an increase in the number of subjects, but the rate of increase in accuracy slows down after a certain point (n=50). Moreover, when the number of subjects is 60, the accuracy of MACTN remains around 55\%. The accuracy of MACTN for 70 subjects is only 0.05\% higher than that for 60 subjects, and the accuracy for 80 subjects is 55.42\%. Thanks to such a large dataset, this work has the potential to establish a high-performance and robust emotion model, thereby expanding the range of applicable populations for emotional brain-computer interfaces.

\subsection{Explainability Analysis}
\begin{figure*}[htbp]
\centering
\includegraphics[width=0.8\textwidth]{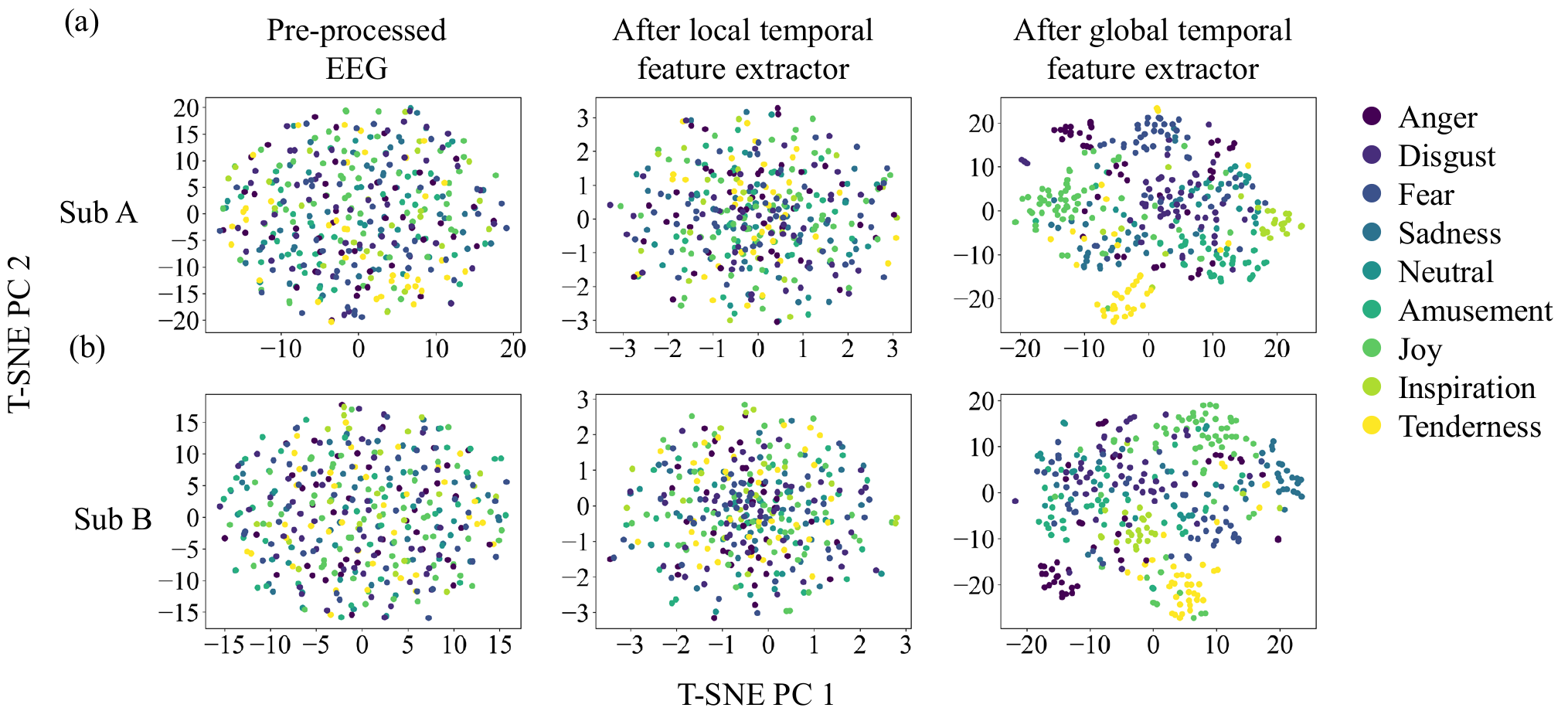}
\caption{Scatter plot of features after dimensionality reduction, using T-SNE method. Each row represents one subject, and Sub A-B represents two typical subjects respectively. Each column represents the features after the calculations of the flow. (a)Top row shows the topological changes of the samples of Sub A after different module processing. (b)Bottom row represents Sub B, and the rest is the same as (a).}
\label{fig_scatter}
\end{figure*}
% \FloatBarrier

\iffalse
para 1 dis
\fi
In order to investigate the topological changes of the samples after being processed by LTFE and GTFE, we selected two parts of the feature maps for visualization, namely the ones that have undergone LTFE and GTFE respectively. We utilized T-SNE to reduce the dimensionality of the feature maps, and displayed them in a scatter plot. In Fig. \ref{fig_scatter}, each point represents a sample, that is, an EEG segment or a feature map obtained through feature extraction. Each row in Figure 4 represents a subject. From left to right, the inter-class and intra-class distances between preprocessed EEG segments are relatively large. After LTFE, the distances between samples in both intra-class and inter-class have been reduced, while after GTFE, the inter-class distances between samples have been enlarged while the intra-class distances remain unchanged.

\begin{figure*}[htbp]
\centering
\includegraphics[width=0.9\textwidth]{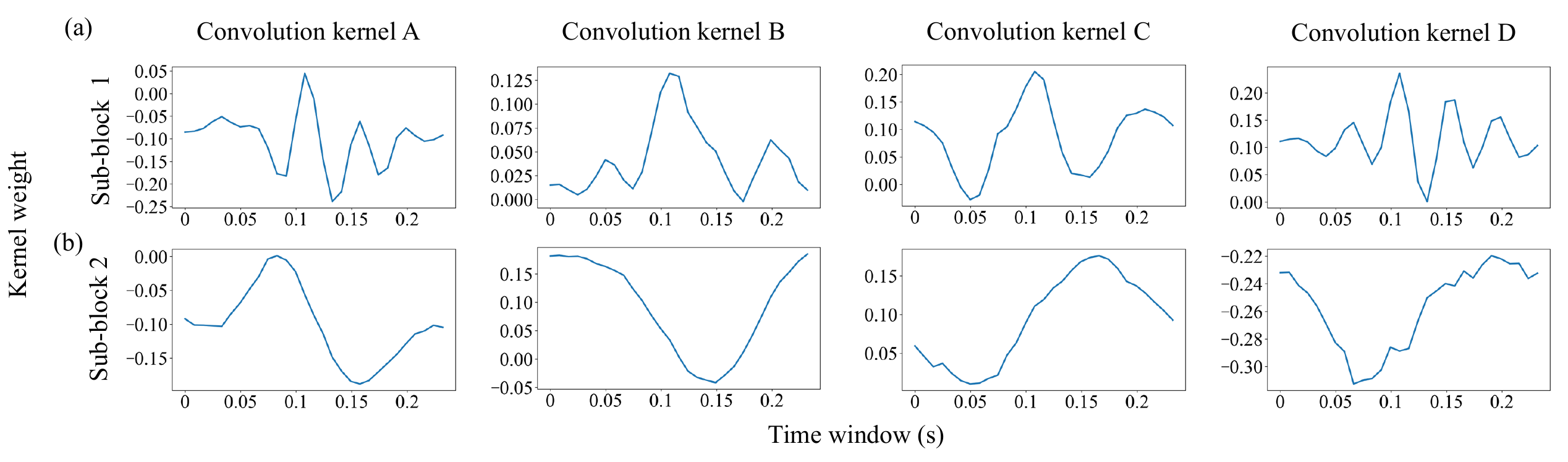}
\caption{Visualization of Separable Conv-Block temporal convolution kernel. (a)Top row represents Separable Conv-Block 1. (a)Bottom row represents Separable Conv-Block 2. Convolution kernel A-D represents several typical feature channels.}
\label{fig_curve}
\end{figure*}

\iffalse
para 2 dis
\fi
Through ablative studies, we found that the Separable Conv-Blocks in LTFE have a significant impact on model accuracy. Therefore, we visualized the convolutional kernels within these sub-blocks. Each subplot in Fig. \ref{fig_curve} displays the learned temporal kernels for a 0.232-second window. The results indicate that the convolutional filters located after the processing stream extract lower frequency information compared to those in the front stream. This finding is consistent with other electrophysiological studies demonstrating that the human brain attends to low-frequency components when processing emotional information, such as those works by LowryK et al. \cite{kirkby2018amygdala} and Huang et al. \cite{huang2021increased}.

% \FloatBarrier

\begin{figure}[!t]
\centering
\includegraphics[width=0.4\textwidth]{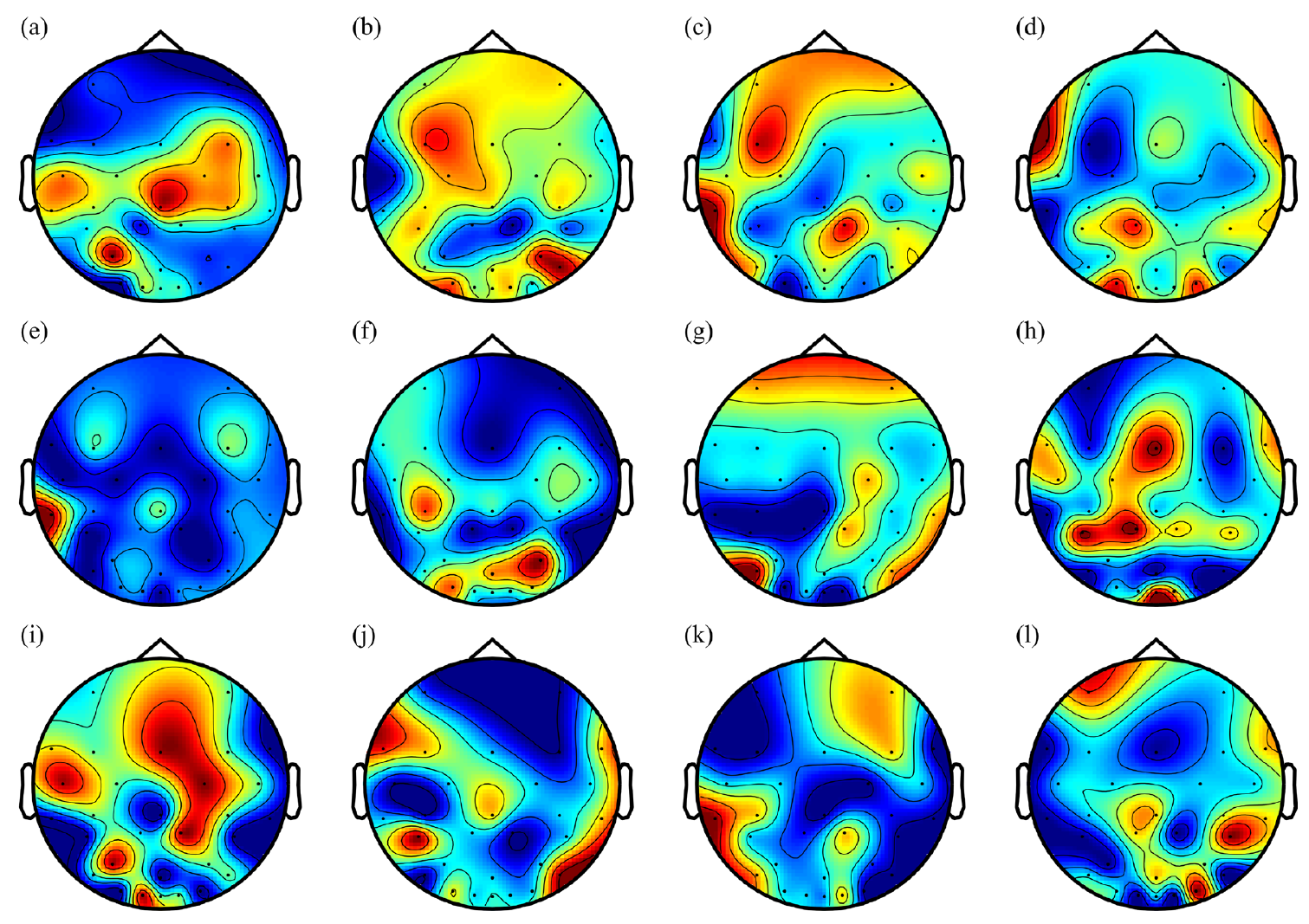}
\caption{Channel attention weight topology map of MACTN in THU-EP. The first row (a)-(d) shows the average topology map of all 10-fold MACTN, where (a)-(d) represent the channel attention weight topology map of feature maps after convolution with different kernel sizes. The second row (e)-(h) shows the topology map of one fold MACTN with the highest accuracy. The third row (i)-(l) shows the topology map of one fold MACTN with the lowest accuracy.}
\label{fig_topo}
\end{figure}

\iffalse
para 3 dis
\fi
In LTFE, we utilized the channel attention sub-block of SK attention, which mainly extracts information of different time and frequency components based on small convolutional kernels of different sizes, and integrates them according to the computed channel attention. Fig. \ref{fig_topo} shows the topological map of the average channel attention weights in the proposed model, which is normalized between -1 and 1 for better visualization. The proposed model uses kernel sizes of 1, 3, 5, and 7, all of which are smaller than those used in previous convolutional networks. The channel attention weights of different kernel sizes can be calculated according to equation (\ref{eq_chanattn}). Fig. \ref{fig_topo} (a)-(d) are the average topological maps of all 10-fold MACTN, Fig. \ref{fig_topo} (e)-(h) are the topologies of the most accurate fold, and Fig. \ref{fig_topo} (i)-(l) are the topologies of the least accurate fold among the 10 folds. Based on Fig. \ref{fig_topo} (a)-(d), the branch with a convolutional kernel size of 1 mainly focuses on information in the parietal region, the branch with a kernel size of 3 pays more attention to information in the parietal and temporal regions, the branch with a kernel size of 5 focuses on information in the frontal and temporal regions, and the branch with a kernel size of 7 primarily attends to information in the temporal and occipital regions. Comparing the topological maps of the most accurate fold and the average topological map, the topological map of the most accurate fold shows a similar focus area to the average topological map.

% \FloatBarrier

\begin{figure}[!ht]
\centering
\includegraphics[width=0.49\textwidth]{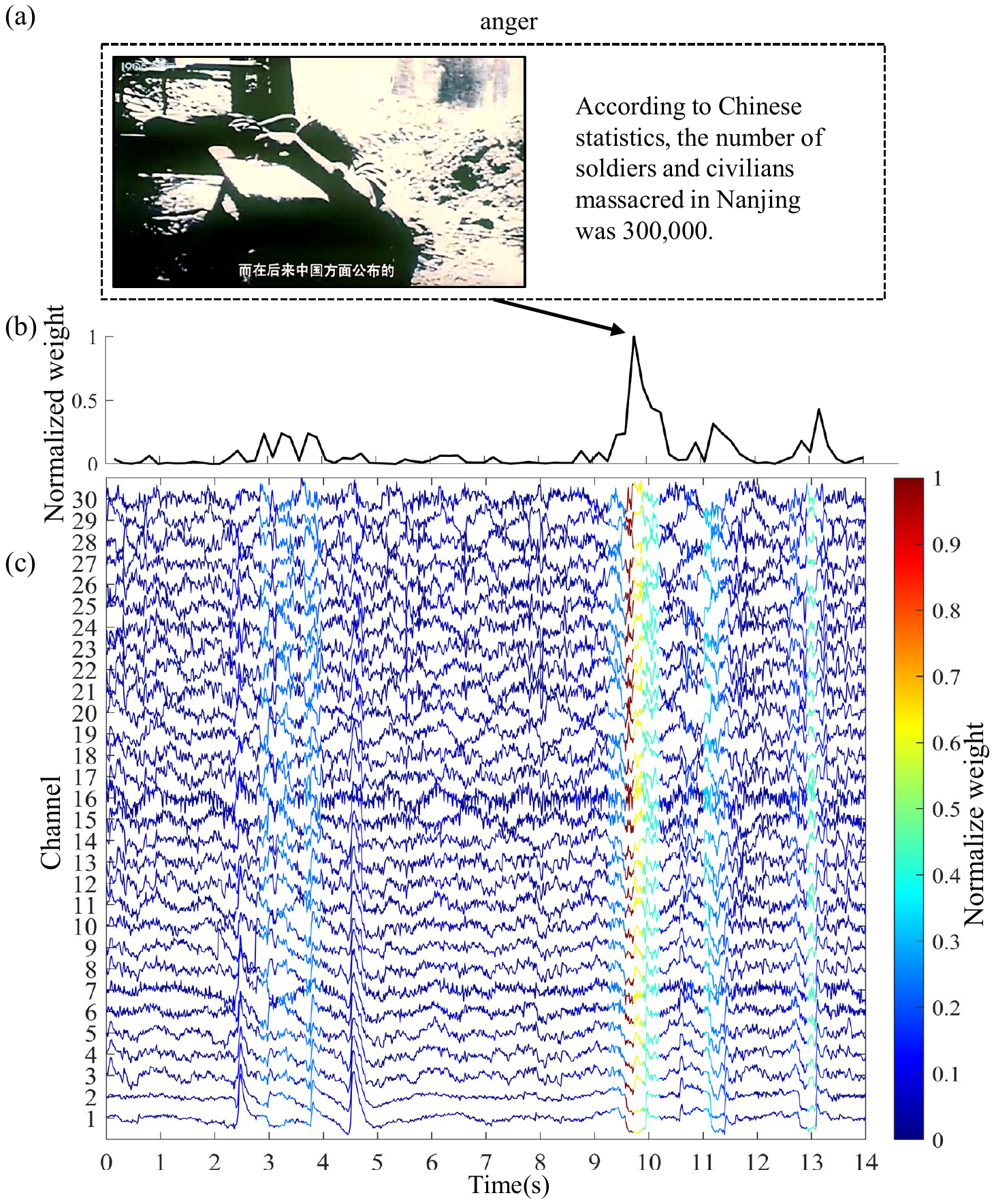}
\caption{Self-attention weight visualization (normalized). This segment is derived from trial 1, which contains the emotion of anger. The time segments with higher self-attention weights correspond to the angry parts of the video segments. (a) Two screenshots from the stimuli and the descriptions corresponding to them respectively. (b) Plot of the normalized self-attention weights. (c) Waveforms of EEG signals, color bar represents normalized self-attention weights.}
\label{fig_sa}
\end{figure}

\iffalse
para 4 dis
\fi
The proposed model treats a feature map of a short period as a word in the natural language processing domain, treats the feature channels as embeddings of each ”word,” and adds an extra learnable class token with sentiment information for each time segment. The attention of each time segment can be calculated based on the equation (\ref{eq_sa}). Fig. \ref{fig_sa} shows an EEG fragment from trial 1. Fig. \ref{fig_sa}(a) shows the screenshot of the stimulus (the stimulus mainly induces anger emotion). Fig. \ref{fig_sa}(b) shows the self-attention weight curve of the EEG fragment (the weight is normalized). Fig. \ref{fig_sa}(c) shows the waveform of this EEG fragment The caption shown in Fig. \ref{fig_sa}(a) is "According to Chinese statistics, the number of soldiers and civilians massacred in Nanjing was 300,000", and Fig. \ref{fig_sa}(a) corresponds to the part of \ref{fig_sa}(b) and \ref{fig_sa}(c) with higher attention weight.

\subsection{Temporal Feature Extractor and Attention Mechanism}
\iffalse
para 7 dis
\fi
The reason why MACTN can achieve high performance is related to the cascaded feature extraction blocks. Consistent with most previous works \cite{shen2022contrastive, liang2021eegfusenet}, we use CNN as a local temporal feature extractor. In many fields, especially in computer vision, CNN has demonstrated its powerful capability in local feature extraction \cite{zhang2018feature}. Research has shown that the intensity of emotions varies over time. We believe that CNN can extract emotional representations with short-term high intensity. Unlike previous works, we use transformer-based attention mechanism to integrate global temporal features, which assigns higher attention weights to segments with higher emotional intensity, thus achieving higher recognition performance. By visualizing the attention weights of self-attention, we found that the parts with high emotional intensity of stimulus-induced materials correspond to high attention weights. Attention weight assignment is sample-specific, which can extract highly variable emotional representations in terms of duration. Sun et al. \cite{sun2022continuous} found that the origin of epileptic seizures was near the epileptic focus by visualizing the self-attention weights, and the electrodes in those regions had higher self-attention weights. In addition to self-attention, SK attention can further improve the model's performance. By visualizing the attention weights of SK attention, it uses convolution with multiple small-size kernels to extract features of different scales and selectively integrates features of different scales (different brain regions) by means of channel attention mechanism. Studies have shown that different brain regions have different selectivity for emotions, such as the amygdala's selectivity for negative emotions \cite{fedele2020relation}, and direct electrical stimulation of the orbitofrontal cortex can improve the emotional state of depressed patients \cite{rao2018direct}.

\subsection{Limitations and Future Work}
\iffalse
此前诸多研究表明基于迁移学习的方法能够提升跨被试情绪识别的性能[MaBQ,2019;LiJ,2020,TCyb；LiR，2021]，域泛化是迁移学习的一种方法，在未来可以使用域泛化进行跨被试情绪识别。
[模型]我们的模型关注了情绪的时间动态性，并在有限的刺激材料上给予了模型一定的解释性。Panagiotis等人[Panagiotis，2012]使用额叶不对称支持的多维定向信息分析方法与不对称指数实现丢弃“情绪相关”信息较少的信号段，只保留有价值的信号段。未来的工作中，希望在刺激材料上赋予不同时间段以相应的情绪强度信息，从而实现在时间上的情绪定位。
EEG包含丰富的时间和空间信息，在这些工作中，我们高度关注时间信息，而越来越多的工作开始关注EEG的空间信息，如使用图卷积神经网络来提取空间相关的情绪特征[Zhao，2022， JNE]。在未来的工作中，希望可以尝试结合如图神经网络这样的关注空间信息的模型，提取更加完整的情绪特征，以提高模型的性能。
\fi
Many previous studies have shown that transfer-learning-based methods can improve the performance of cross-subject emotion recognition\cite{ma2019reducing,li2019domain,li2021multi}, and domain generalization is a method of transfer learning that can be used for cross-subject emotion recognition in the future.
Panagiotis et al. \cite{petrantonakis2012adaptive} used frontal asymmetry-supported multidimensional directional information analysis with asymmetric indices to discard signal segments with little "emotion-related" information and retain only valuable signal segments. Future work is expected to be more effective in stimulus analysis. In future work, it is hoped that temporal emotion localization can be achieved by assigning different temporal segments with corresponding emotional intensity information on the stimulus material.
EEG contains rich temporal and spatial information, and we highly focus on temporal information in these works. In contrast, more and more works have started to focus on the spatial information of EEG, such as using graph convolutional neural networks to extract spatially relevant emotional features \cite{zhao2022scc}. In future work, we hope to combine models that focus on spatial information, such as graph neural networks to extract more complete emotional features to improve the model's performance.

\section{Conclusions}
\iffalse
在这项研究中，我们提出了一种基于CNN与transformer混合的模型，该方法受到神经科学中有关情绪的时间动态性研究的启发。该方法使用CNN提取时间上情绪强度较高的局部特征，借助transformer中自注意力机制整合时间上稀疏的全局特征，结合通道注意力机制关注通道维度的特征。值得一提的是该方法是一个端到端的方法，无需手工提取特征。使用该方法，获得了在THU-EP数据集上最好的分类精度，并且MACTN的相关变体获得了2022年世界机器人大赛——脑控机器人大赛情绪赛题的冠军。此外，我们使用了4种可视化方法对模型做出来解释，其中通过可视化重要的注意力权重，我们展现了它在时间维度上进行情绪定位方面的潜力。
\fi
In this study, we propose a model based on a hybrid of CNN and transformer, inspired by research on the temporal dynamics of emotions in neuroscience. The method uses CNN to extract local features with high emotional intensity over time, integrates sparse global features over time using the self-attention mechanism in transformer, and combines channel attention mechanism to focus on features in the channel dimension. It is worth noting that this method is an end-to-end approach that does not require manual feature extraction. Extensive experiments on two public datasets, THU-EP and DEAP, show that in most experimental settings, MACTN achieves higher classification accuracy and F1 scores than other methods. Moreover, an earlier version of the method with the same concept won the championship of the 2022 World Robot Contest – Emotional BCI. In addition, we use four visualization methods to provide insights into the model, and demonstrate its potential in emotion localization over time by visualizing important attention weights.

% https://www.overleaf.com/project/6371cb2a3729367c72366bcb
% or


% \appendices

% \section{}

\ifCLASSOPTIONcompsoc
  \section*{Acknowledgments}
\else
  \section*{Acknowledgment}
\fi

This work was supported by National Natural Science Foundation of China (No. 61906132 and No. 81925020), National Key Research and Development Program of China ‘Biology and Information Fusion’ Key Project (No. 2021YFF1200600), and Key Project \& Team Program of Tianjin City (No. XC202020).

\ifCLASSOPTIONcaptionsoff
  \newpage
\fi



%\bibliographystyle{IEEEtran}
\bibliographystyle{IEEEtran}
\bibliography{references}


% \end{thebibliography}


\begin{IEEEbiography}[{\includegraphics[width=0.96in,height=1.2in,clip,keepaspectratio]{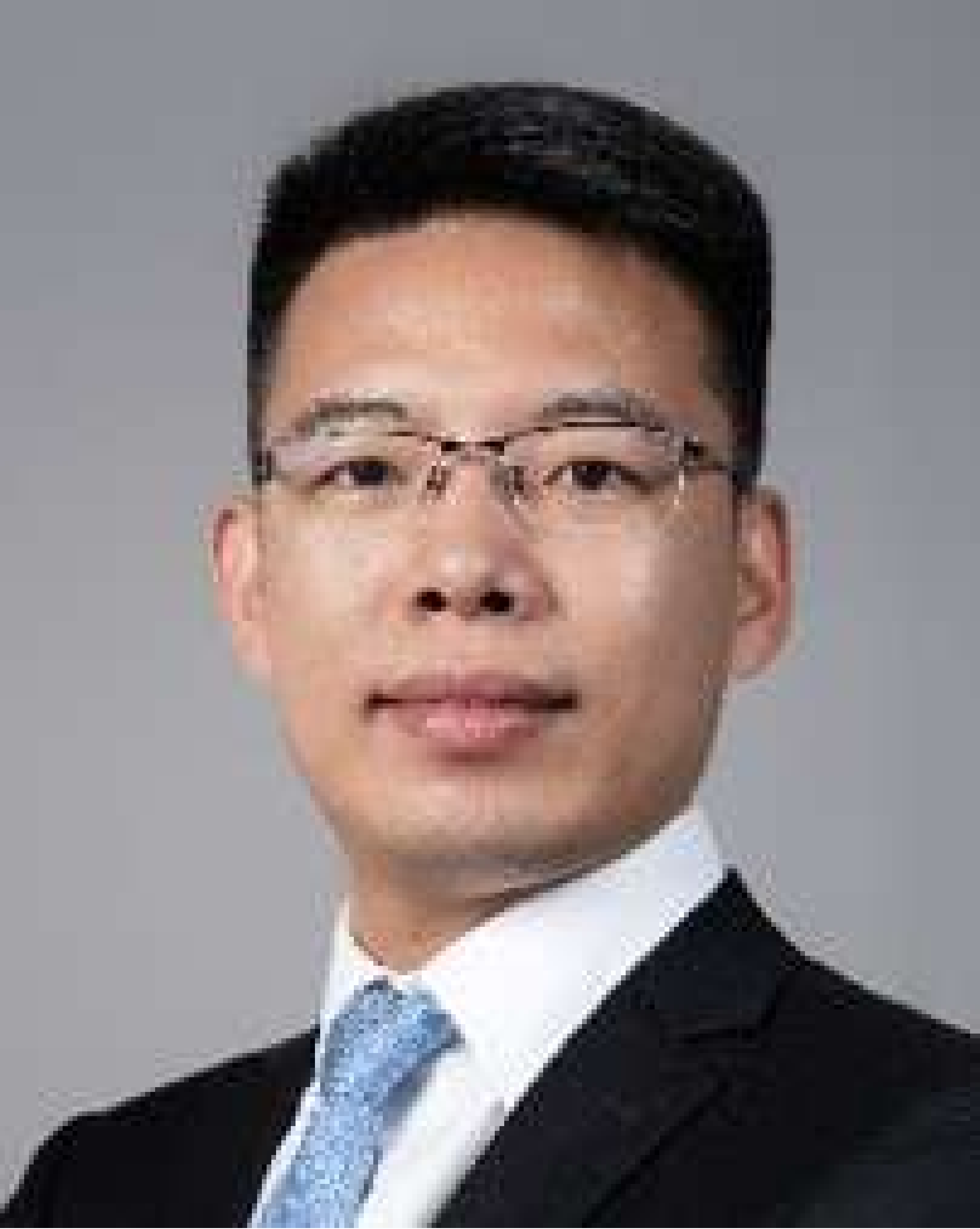}}]{Xiaopeng Si}
(Member, IEEE) works as an associate Professor at Tianjin University, since 2018. He obtained Ph.D. in biomedical engineering from Tsinghua University, and finished the visiting scholar in biomedical engineering department at Johns Hopkins University. His research interest includes cognitive neuroscience and brain-inspired intelligence, neural engineering and brain-computer interaction, neural information acquisition and intelligent computing, multimodal human brain neuroimaging and regulation, etc. His team used multimodal neuroimaging and machine learning methods to understand the human cognitive processes. They tried to apply these mechanisms in people with brain disorders, and also tried to enhance normal people's ability by creating mind reading technology. Dr. Si has managed and participated many national and international research projects. He has published papers in PNAS, Cerebral Cortex, IEEE JBHI, JNE, Frontiers in Aging Neuroscience, Frontiers in Neuroscience, etc. 
His team won the first prize of Emotional BCI Competition during 2022 World Robot Contest, and the second prize of BCI Competition during 2020 World Robot Contest.
\end{IEEEbiography}

\begin{IEEEbiography}[{\includegraphics[width=0.96in,height=1.2in,clip,keepaspectratio]{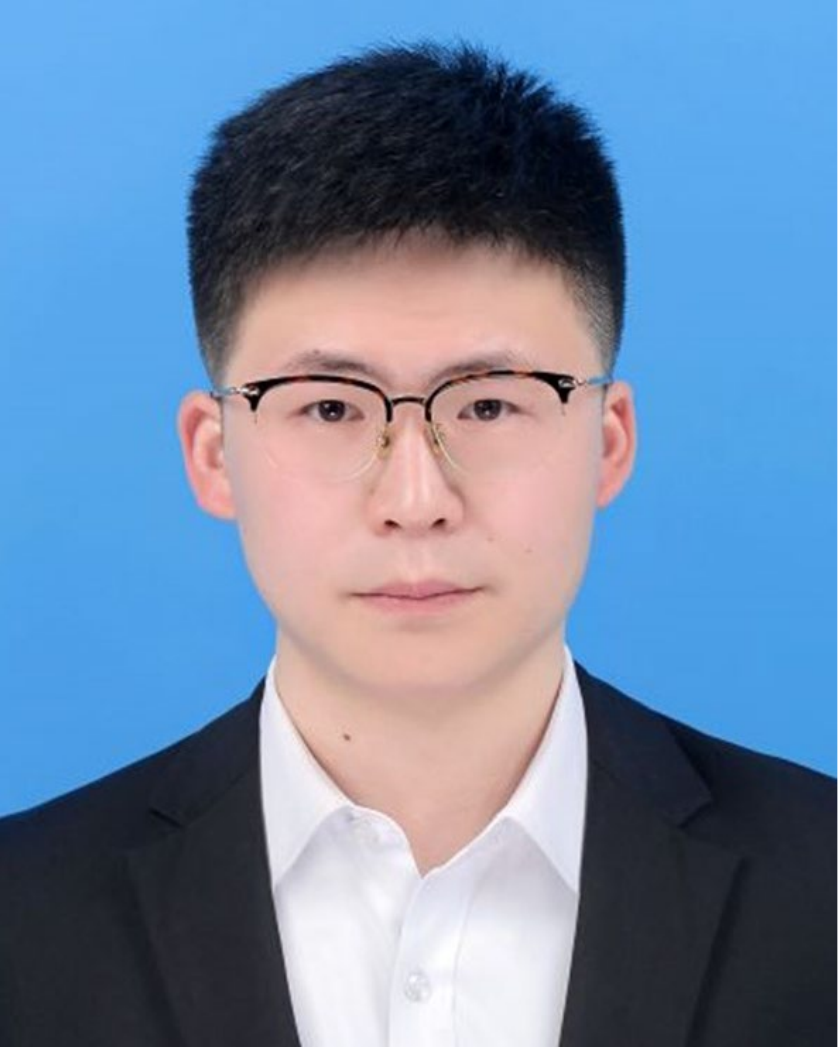}}]{Dong Huang}
received the B.S. degree in electronics information science and technology from Northeast Petroleum University in 2020. He is currently pursuing the M.E. degree in biomedical engineering, Tianjin University, Tianjin, China. His research interest includes affective computing, and deep learning.
\end{IEEEbiography}

 % \vspace{-200 pt}

\begin{IEEEbiography}[{\includegraphics[width=0.96in,height=1.2in,clip,keepaspectratio]{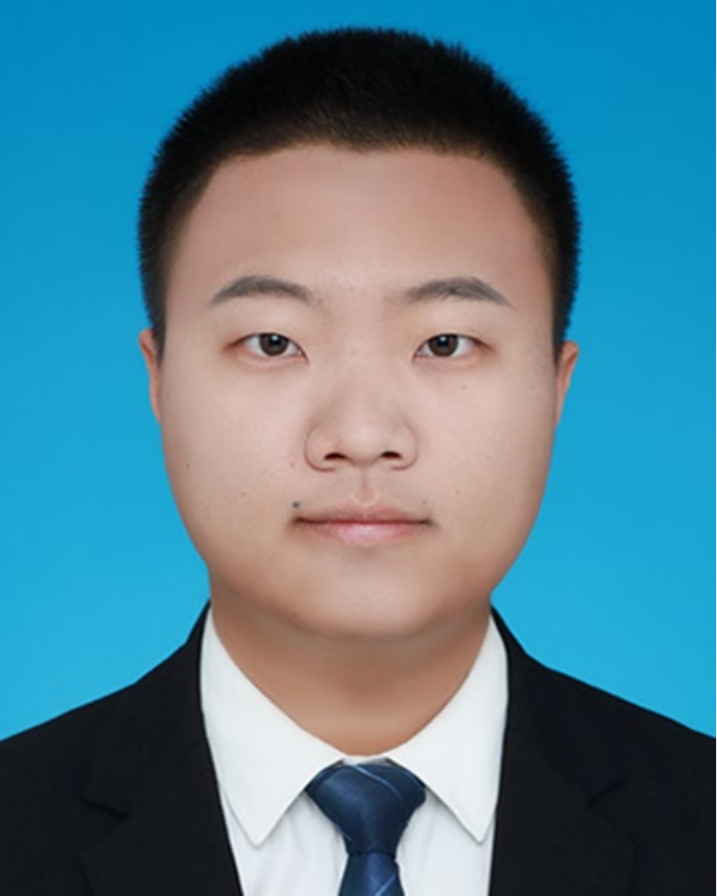}}]{Yuling Sun}
received the B.E. degree in biomedical engineering from Tianjin University in 2020. Currently, he is working toward the M.M. degree in biomedical engineering, Tianjin University, Tianjin, China. His research interest includes seizure detection, digital signal processing, affective computing, and deep learning.
\end{IEEEbiography}

 % \vspace{-200 pt}

\begin{IEEEbiography}[{\includegraphics[width=0.96in,height=1.2in,clip,keepaspectratio]{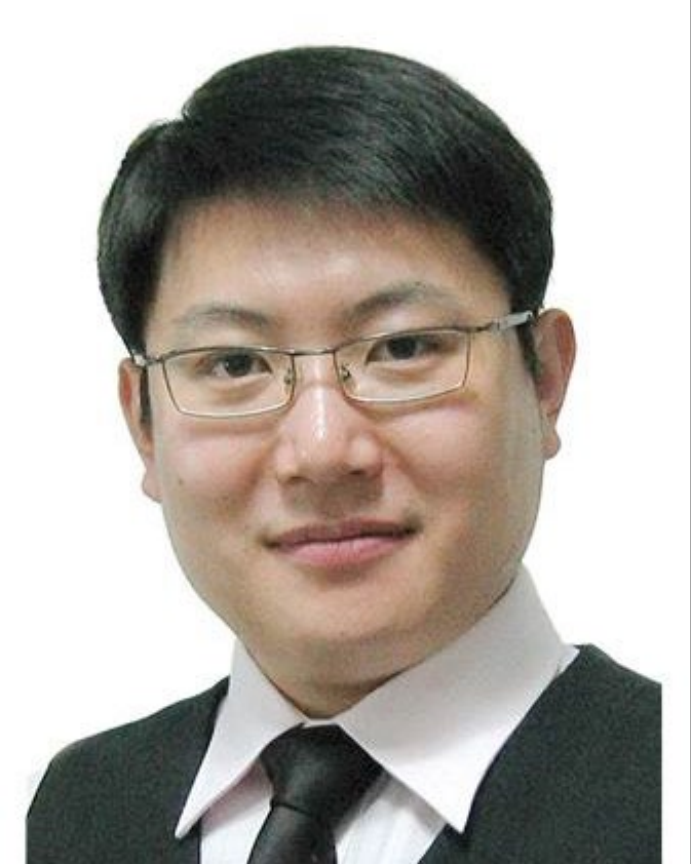}}]{Dong Ming}
(Senior Member, IEEE) received the B.S. and Ph.D. degrees in biomedical engineering from Tianjin University, Tianjin, China, in 1999 and 2004, respectively. He worked as a Research Associate at the Department of Orthopaedics and Traumatology, Li Ka Shing Faculty of Medicine, The University of Hong Kong, from 2002 to 2003, and was a Visiting Scholar with the Division of Mechanical Engineering and Mechatronics, University of Dundee, U.K., from 2005 to 2006. He joined Tianjin University (TJU), as a Faculty at the College of Precision Instruments and Optoelectronics Engineering, in 2006, and has been promoted to a Full Professor of biomedical engineering, since 2011. He is currently a Chair Professor with the Department of Biomedical Engineering, TJU, where he is also the Head of the Neural Engineering and Rehabilitation Laboratory. His major research interests include neural engineering, rehabilitation engineering, sports science, biomedical instrumentation, and signal/image processing, especially in functional electrical stimulation, gait analysis, and brain–computer interface. Furthermore, he has been an International Advisory Board Member of the The Foot, and the Editorial Committee Member of Acta Laser Biology Sinica, and International Journal of Biomedical Engineering, China. He has managed over ten national and international research projects, organized and hosted several international conferences, as the Session Chair or Track Chair over the last ten years and was the General Chair of the 2012 IEEE International Conference on Virtual Environments, Human–Computer Interfaces and Measurement Systems (VECIMS 12). He is the Chair of the IEEE-EMBS Tianjin Chapter.

% \vspace{200 pt}

\end{IEEEbiography}


%\vfill


\end{document}